\documentclass[prl, twocolumn,superscriptaddress, nofootinbib, longbibliography]{revtex4-2}

\usepackage[T1]{fontenc}
\usepackage[utf8]{inputenc}
\usepackage{color}
\usepackage{amsmath}
\usepackage{amssymb}
\usepackage{graphicx}
\usepackage{wasysym}
\usepackage[pdfusetitle,
 bookmarks=true,bookmarksnumbered=false,bookmarksopen=false,
 breaklinks=false,pdfborder={0 0 0},pdfborderstyle={},backref=false,colorlinks=true]
 {hyperref}
\hypersetup{
 citecolor=blue,urlcolor=blue}



\begin{document}
\title{Reducing thermal noises by quantum refrigerators }
\author{Han-Jia Bi}
\author{Sheng-Wen Li}
\email{lishengwen@bit.edu.cn}

\affiliation{Center for Quantum Technology Research, and Key Laboratory of Advanced
Optoelectronic Quantum Architecture and Measurements, School of Physics,
Beijing Institute of Technology, Beijing 100081, People’s Republic
of China}
\begin{abstract}
Reducing the thermal noises in microwave (MW) resonators can bring
about significant progress in many research fields. In this study,
we consider using three-level or four-level systems as “quantum refrigerators”
to cool down MW resonators so as to reduce the thermal noises, and
investigate their possible cooling limits. In such a quantum refrigerator
system, the MW resonator is coupled with many three-level or four-level
systems. Proper light pump makes the multilevel systems concentrated
into their ground states, which continuously absorb the thermal photons
in the MW resonator. By adiabatic elimination, we give a more precise
description for this cooling process. For three level systems, though
the laser driving can cool down the multilevel systems efficiently,
a too strong driving strength also significantly perturbs their energy
levels, breaking the resonant interaction between the atom and the
resonator, which weakens the cooling effect, and that sets a finite
region for cooling parameters. In four level systems, by adopting
an indirect pumping approach, such a finite cooling region can be
further released. In both cases, we obtain analytical results for
the cooling limit of the MW resonator. Based on practical parameters,
our estimation shows the cooling limit could reach lower than the
liquid helium temperature, without resorting to the traditional cryogenic
systems.
\end{abstract}
\maketitle

\section{Introduction }

Microwave (MW) resonators are essential electronic devices widely
used in many research areas, such as the signal radiation and detection
in communication systems \citep{yuen_deep_1983}, cosmology radio
telescopes \citep{wilson_tools_2013}, and electron/nuclear spin resonance
spectrometers \citep{lund_principles_2011,gunther_nmr_2013}. But
at room temperature ($T\sim300\,\text{K}$), the thermal noises in
MW resonators are generally quite strong. For the instance of an MW
resonator with $\omega_{\text{\textsc{r}}}/2\pi=1\,\text{GHz}$, the
thermal photon number in the MW resonator is $\sim6\times10^{3}$.
The signals weaker than this intensity level would be buried in the
fluctuating noise background. Thus, generally a complicated cryogenic
system is needed to cooled down the system to the liquid helium temperature
($T\simeq4\,\text{K}$, thermal photon number $\apprle10^{2}$) \citep{siegman_microwave_1964,oxborrow_room-temperature_2012,jin_proposal_2015,wu_enhanced_2022}. 

Alternatively, a bench-top cooling method by ``quantum refrigerators''
is adopted to cool down the MW resonators \citep{wu_bench-top_2021,blank_anti-maser_2023,chen_overcoming_2024,ng_quasi-continuous_2021,fahey_steady-state_2023,gottscholl_room-temperature_2023,wang_spin-refrigerated_2024,day_room-temperature_2024},
where the MW resonator is coupled with an ensemble of multilevel systems.
By proper light pump \citep{ng_quasi-continuous_2021,gottscholl_room-temperature_2023,fahey_steady-state_2023,day_room-temperature_2024,wang_spin-refrigerated_2024},
or MW radiation \citep{wu_bench-top_2021,chen_overcoming_2024,blank_anti-maser_2023},
the ensemble populations can be concentrated into the ground states,
then the thermal photons in the MW resonator can be continuously absorbed
by the ensemble, where the heat is dumped away through light radiation.
It is reported that the liquid nitrogen temperature can be reached
($\sim66\,\text{K}$ \citep{day_room-temperature_2024}). However,
because of the complicated noise environments in solid defects systems,
some theoretical analysis also shows this is almost the cooling limit
of such system \citep{zhang_microwave_2022}. To get a better cooling
effect, some more improvements are still needed. 

It is worth noting that such systems are quite similar as the Scovil--Schulz-DuBois--Geusic
(SSDG) quantum refrigerator \citep{scovil_three-level_1959,geusic_quantum_1967,boukobza_thermodynamic_2006,boukobza_three-level_2007,scully_quantum_2011,kosloff_quantum_2013,uzdin_equivalence_2015,wang_four-level_2015,li_quantum_2017,cao_quantum_2022}.
Here, we make a thorough analysis on the possible cooling limit of
the MW resonator, by adopting more simplified three-level or four-level
structures as the quantum refrigerator. The lowest two atom levels
are resonantly coupled with the resonator mode, and a driving laser
is applied on the atom, which could make the atom population fully
concentrated into the ground state, effectively giving a zero temperature.
By adiabatic elimination \citep{cirac_laser_1992,scully_quantum_1997,gardiner_quantum_2004},
we derive a master equation for the resonator mode, which gives a
more precise description. 

We find that, for three level systems, the atom can be cooled down
quite efficiently, but a too strong driving strength also brings in
significant perturbation to the atom levels. Such perturbation could
break the resonant interaction between the atom and the resonator,
which prevents the heat transport from the MW resonator to the refrigerator,
and weakens the cooling effect. That sets a finite region for cooling
parameters. By utilizing an indirect pumping approach in a four-level
refrigerator \citep{wang_four-level_2015,li_quantum_2017,yan_external-level_2021},
such a constraint condition can be released. In this case, the driving
laser is applied on the upper two levels, thus no longer perturb the
resonant coupling between the lower two levels and the resonator,
and the heat can be indirectly absorbed away like the ``siphonic''
effect. In both cases, we obtain analytical results for the cooling
limit of the MW resonator. Based on some practical parameters in current
experiments, our estimation shows the cooling limit of the MW resonator
could reach the liquid helium temperature. 

\section{Quantum refrigerator setup}

\begin{figure}
\begin{centering}
\includegraphics[width=0.95\columnwidth]{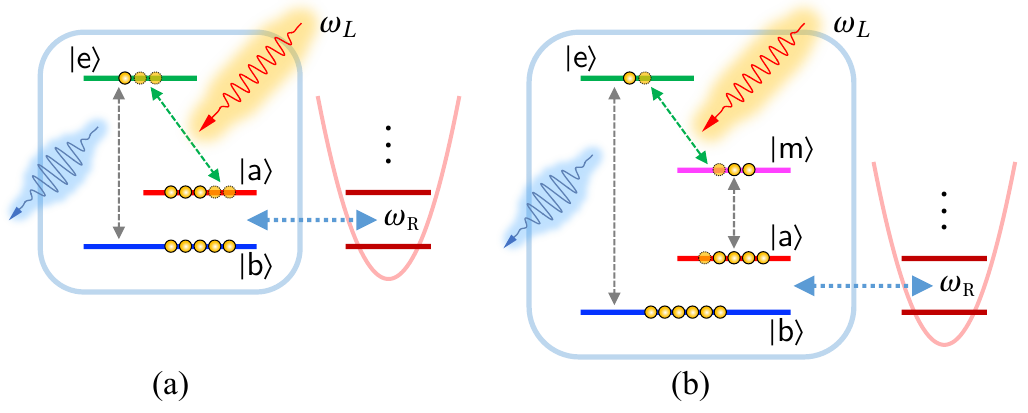}
\par\end{centering}
\caption{Demonstration for the interaction between a MW resonator mode and
(a) a three-level atom, (b) a four-level atom. Here $\omega_{L}$
is the frequency of the driving laser, and $\omega_{\text{\textsc{r}}}$
is the frequency of the MW mode. The transition pathways are indicated
by the dashed lines. In the three-level atom, the laser driving could
bring in perturbation to $|\mathsf{a}\rangle$, which could break
the exchanging resonance between $|\mathsf{a,b}\rangle$ and the MW
resonator. }

\label{fig-demo}
\end{figure}

In analog to the SSDG quantum refrigerator \citep{scovil_three-level_1959,geusic_quantum_1967,boukobza_thermodynamic_2006,boukobza_three-level_2007,scully_quantum_2011,kosloff_quantum_2013,uzdin_equivalence_2015,wang_four-level_2015,li_quantum_2017,cao_quantum_2022},
here we consider using a three-level atom as a quantum refrigerator
to cool down the MW resonator ($\hat{H}_{\text{\textsc{r}}}=\omega_{\text{\textsc{r}}}\,\hat{a}^{\dag}\hat{a}$).
The self Hamiltonian of the atom is described by $\hat{H}_{\text{\textsc{a}}}=\sum_{\alpha}\,\varepsilon_{\alpha}|\alpha\rangle\langle\alpha|$
{[}$\alpha=\mathsf{a,b,e}$, see Fig.\,\ref{fig-demo}(a){]}. The
energy gap ($\Omega_{\mathsf{ab}}:=\varepsilon_{\mathsf{a}}-\varepsilon_{\mathsf{b}}$)
between the lowest two levels $|\mathsf{a}\rangle$, $|\mathsf{b}\rangle$
is in resonance with the MW resonator $\omega_{\text{\textsc{r}}}$,
and their interaction is described by $\hat{V}_{\text{\textsc{ar}}}=g\,(\hat{\sigma}^{+}\hat{a}+\hat{\sigma}^{-}\hat{a}^{\dag})$.
Here, we denote\footnote{In this paper, generally we denote $\hat{\tau}_{\alpha\beta}^{+}:=|\alpha\rangle\langle\beta|:=(\hat{\tau}_{\alpha\beta}^{-})^{\dag}$
as the transition operator between $|\alpha\rangle$, $|\beta\rangle$
(for $\varepsilon_{\alpha}>\varepsilon_{\beta}$), and $\Omega_{\alpha\beta}:=\varepsilon_{\alpha}-\varepsilon_{\beta}$
as the energy gap. The MW transition operator $\hat{\sigma}^{\pm}$
is equivalent with $\hat{\tau}_{\mathsf{ab}}^{\pm}$.} $\hat{\sigma}^{+}:=|\mathsf{a}\rangle\langle\mathsf{b}|=(\hat{\sigma}^{-})^{\dag}$,
and $g$ is the atom-resonator coupling strength. The energy gaps
$\Omega_{\mathsf{ea}}$, $\Omega_{\mathsf{eb}}$ are in the optical
regime. 

In addition, a driving laser with frequency $\omega_{L}=\Omega_{\mathsf{ea}}$
is resonantly applied to the transition $|\mathsf{e}\rangle\leftrightarrow|\mathsf{a}\rangle$,
which is described by $\hat{V}_{\text{d}}(t)=\tilde{\Omega}_{\text{d}}\,(\hat{\tau}_{\mathsf{ea}}^{+}+\hat{\tau}_{\mathsf{ea}}^{-})\cos\omega_{L}t\simeq\frac{1}{2}\tilde{\Omega}_{\text{d}}\,(\hat{\tau}_{\mathsf{ea}}^{+}e^{-i\omega_{L}t}+\mathrm{H.c.})$.
Here $\tilde{\Omega}_{\text{d}}$ is the Rabi frequency, which characterizes
the driving light intensity.

The dynamics of the composite atom-resonator state $\rho(t)$ is described
by the following master equation (interaction picture) 
\begin{align}
\partial_{t}\rho & =i[\rho,\,\tilde{V}_{\text{d}}+\tilde{V}_{\text{\textsc{ar}}}]+\mathcal{D}_{\text{\textsc{a}}}[\rho]+\mathcal{D}_{\text{dep}}[\rho]+\mathcal{D}_{\text{\textsc{r}}}[\rho],\\
\mathcal{D}_{\text{\textsc{r}}}[\rho] & =\kappa\bar{\mathsf{n}}_{\text{\textsc{r}}}\big(\hat{a}^{\dag}\rho\hat{a}-\tfrac{1}{2}\{\hat{a}\hat{a}^{\dag},\,\rho\}\big)\nonumber \\
 & +\kappa(\bar{\mathsf{n}}_{\text{\textsc{r}}}+1)\big(\hat{a}\rho\hat{a}^{\dag}-\tfrac{1}{2}\{\hat{a}^{\dag}\hat{a},\,\rho\}\big),\nonumber \\
\mathcal{D}_{\text{\textsc{a}}}[\rho] & =\sum_{\alpha,\beta}^{\varepsilon_{\alpha}>\varepsilon_{\beta}}\Gamma_{\alpha\beta}^{+}\big(\hat{\tau}_{\alpha\beta}^{+}\rho\hat{\tau}_{\alpha\beta}^{-}-\tfrac{1}{2}\{\hat{\tau}_{\alpha\beta}^{-}\hat{\tau}_{\alpha\beta}^{+},\rho\}\big)\nonumber \\
 & +\Gamma_{\alpha\beta}^{-}\big(\hat{\tau}_{\alpha\beta}^{-}\rho\hat{\tau}_{\alpha\beta}^{+}-\tfrac{1}{2}\{\hat{\tau}_{\alpha\beta}^{+}\hat{\tau}_{\alpha\beta}^{-},\rho\}\big),\nonumber \\
\mathcal{D}_{\text{dep}}[\rho] & =\sum_{\alpha}\tilde{\gamma}_{\text{p}}^{\alpha}\big(\hat{\text{\textsc{n}}}_{\alpha}\rho\hat{\text{\textsc{n}}}_{\alpha}-\tfrac{1}{2}\{\hat{\text{\textsc{n}}}_{\alpha},\rho\}\big).\nonumber 
\end{align}
 Here, $\mathcal{D}_{\text{\textsc{r}}}[\rho]$ and $\mathcal{D}_{\text{\textsc{a}}}[\rho]$
describe the dissipation effect of the MW mode and the atom respectively.
We denote $\kappa$ as the resonator decay rate, $\Gamma_{\alpha\beta}^{+}:=\gamma_{\alpha\beta}\bar{\mathsf{n}}_{\alpha\beta}$,
$\Gamma_{\alpha\beta}^{-}:=\gamma_{\alpha\beta}(\bar{\mathsf{n}}_{\alpha\beta}+1)$
as the dissipation rates of the atom, where $\gamma_{\alpha\beta}$
are the spontaneous decay rates, and $\bar{\mathsf{n}}_{\text{\textsc{r}}}:=(e^{\omega_{\text{\textsc{r}}}/T}-1)^{-1}$,
$\bar{\mathsf{n}}_{\alpha\beta}:=(e^{\Omega_{\alpha\beta}/T}-1)^{-1}$
are the Planck functions with temperature $T$. These dissipation
rates satisfy the Boltzmann ratio $\Gamma_{\alpha\beta}^{+}/\Gamma_{\alpha\beta}^{-}=e^{-\Omega_{\alpha\beta}/T}$. 

In the atom dissipation term $\mathcal{D}_{\text{\textsc{a}}}[\rho]$
we only consider the optical transitions $|\mathsf{e}\rangle\leftrightarrow|\mathsf{a}\rangle$,
$|\mathsf{e}\rangle\leftrightarrow|\mathsf{b}\rangle$ {[}the dashed
paths in Fig.\,\ref{fig-demo}(a){]}. Since $\Omega_{\mathsf{ab}}$
is in the MW regime, generally the spontaneous decay rate between
$|\mathsf{a}\rangle$, $|\mathsf{b}\rangle$ is negligibly small.
In addition, the pure dephasing effects of the atom levels are also
taken into account, which is described by $\mathcal{D}_{\text{dep}}[\rho]$
(here, $\hat{\text{\textsc{n}}}_{\alpha}:=|\alpha\rangle\langle\alpha|$,
and $\tilde{\gamma}_{\text{p}}^{\alpha}$ are the pure dephasing rates). 

Such a three level system can be regarded as an SSDG quantum refrigerator
\citep{scovil_three-level_1959,geusic_quantum_1967,wang_four-level_2015,cao_quantum_2022}.
A driving laser is applied on $|\mathsf{a}\rangle\leftrightarrow|\mathsf{e}\rangle$,
making the population on $|\mathsf{a}\rangle$ greatly reduced and
approach zero, and then the ``heat'' is dumped away through the
optical emission $|\mathsf{e}\rangle\rightarrow|\mathsf{b}\rangle$.
Effectively, that makes the three-level atom become a system with
zero temperature, which could absorb ``heat'' from the MW resonator.
By this way the thermal photons in the MW resonator can be continuously
reduced by the quantum refrigerator. 

\section{Photon dynamics of the MW mode }

To give a more precise description for the above cooling process,
we need a dynamical equation solely for the resonator state $\varrho_{\text{\textsc{r}}}\equiv\mathrm{tr}_{\text{\textsc{a}}}\rho$.
Generally speaking, the atom could achieve its steady state much faster
than the resonator mode. That enables us to apply the adiabatic elimination
\citep{cirac_laser_1992,scully_quantum_1997,gardiner_quantum_2004},
which finally gives an equation for the MW mode alone, that is (see
Appendix A), 
\begin{align}
\partial_{t}\varrho_{\text{\textsc{r}}}= & (A_{+}+\kappa\bar{\mathsf{n}}_{\text{\textsc{r}}})\big(\hat{a}^{\dagger}\varrho_{\text{\textsc{r}}}\hat{a}-\tfrac{1}{2}\{\varrho_{\text{\textsc{r}}},\,\hat{a}^{\dagger}\hat{a}\}\big)\nonumber \\
 & +[A_{-}+\kappa(\bar{\mathsf{n}}_{\text{\textsc{r}}}+1)]\big(\hat{a}\varrho_{\text{\textsc{r}}}\hat{a}^{\dag}-\tfrac{1}{2}\{\varrho_{\text{\textsc{r}}},\,\hat{a}\hat{a}^{\dagger}\}\big).\label{eq:mu-t}
\end{align}
 Here, $A_{-(+)}$ can be regarded as the cooling (heating) rate induced
by the atom, and they are given by 
\begin{equation}
A_{\pm}:=2g^{2}\,\mathrm{Re}\int_{0}^{\infty}ds\,\langle\hat{\sigma}^{\pm}(s)\hat{\sigma}^{\mp}(0)\rangle_{\text{ss}}.\label{eq:A}
\end{equation}
 Here $\langle\hat{\sigma}^{\pm}(s)\hat{\sigma}^{\mp}(0)\rangle_{\text{ss}}$
is the time correlation function of the atom in the steady state when
it is not coupled with the MW resonator, which is described by ($\varrho_{\text{\textsc{a}}}\equiv\mathrm{tr}_{\text{\textsc{r}}}\rho$
is the atom state) 
\begin{equation}
\partial_{t}\varrho_{\text{\textsc{a}}}=i[\varrho_{\text{\textsc{a}}},\,\tilde{V}_{\text{d}}]+\mathcal{D}_{\text{\textsc{a}}}[\varrho_{\text{\textsc{a}}}]+\mathcal{D}_{\text{dep}}[\varrho_{\text{\textsc{a}}}].\label{eq:atom-ME}
\end{equation}

In the steady state $t\rightarrow\infty$, the above resonator equation
(\ref{eq:mu-t}) gives the MW photon number as 
\begin{equation}
\langle\hat{n}\rangle_{\text{ss}}=\frac{A_{+}+\kappa\bar{\mathsf{n}}_{\text{\textsc{r}}}}{A_{-}-A_{+}+\kappa}.\label{eq:nss}
\end{equation}
 If the cooling rate is fast enough ($A_{-}\gg A_{+},\,\kappa\bar{\mathsf{n}}_{\text{\textsc{r}}}$),
we obtain $\langle\hat{n}\rangle_{\text{ss}}\rightarrow0$, which
means the thermal noise in the MW resonator is greatly suppressed. 

The time correlation functions $\langle\hat{\sigma}^{\pm}(s)\hat{\sigma}^{\mp}(0)\rangle_{\text{ss}}$
in $A_{\pm}$ {[}Eq.\,(\ref{eq:A}){]} can be calculated with the
help of the quantum regression theorem from the atom equation (\ref{eq:atom-ME})
\citep{scully_quantum_1997,gardiner_quantum_2004,agarwal_quantum_2012,breuer_theory_2002}.
For the three level system {[}Fig.\,\ref{fig-demo}(a){]}, that gives
the heating and cooling rates (\ref{eq:A}) as (the full results are
presented in Appendix B) 
\begin{align}
A_{+} & =\frac{2g^{2}}{\tilde{\Upsilon}_{\mathsf{ab}}+\tilde{\Omega}_{\text{d}}^{2}/4\tilde{\Upsilon}_{\mathsf{eb}}}\mathrm{Re}\big[\langle\hat{\text{\textsc{n}}}_{\mathsf{a}}\rangle_{\text{ss}}+\frac{i\tilde{\Omega}_{\text{d}}}{2\tilde{\Upsilon}_{\mathsf{eb}}}\langle\hat{\tau}_{\mathsf{ea}}^{+}\rangle_{\text{ss}}\big],\nonumber \\
A_{-} & =\frac{2g^{2}}{\tilde{\Upsilon}_{\mathsf{ab}}+\tilde{\Omega}_{\text{d}}^{2}/4\tilde{\Upsilon}_{\mathsf{eb}}}\langle\hat{\text{\textsc{n}}}_{\mathsf{b}}\rangle_{\text{ss}}.\label{eq:A+-3}
\end{align}
 Here, $\tilde{\Upsilon}_{\mathsf{ab}}:=\frac{1}{2}(\Gamma_{\mathsf{ea}}^{+}+\Gamma_{\mathsf{eb}}^{+}+\tilde{\gamma}_{\text{p}}^{\mathsf{a}}+\tilde{\gamma}_{\text{p}}^{\mathsf{b}})$
and $\tilde{\Upsilon}_{\mathsf{eb}}:=\frac{1}{2}(\Gamma_{\mathsf{ea}}^{-}+\Gamma_{\mathsf{eb}}^{-}+\Gamma_{\mathsf{eb}}^{+}+\tilde{\gamma}_{\text{p}}^{\mathsf{e}}+\tilde{\gamma}_{\text{p}}^{\mathsf{b}})$
are the total dephasing rates for the oscillations between the levels
$|\mathsf{a}\rangle\leftrightarrow|\mathsf{b}\rangle$ and $|\mathsf{e}\rangle\leftrightarrow|\mathsf{b}\rangle$
respectively\footnote{When there is no the driving field ($\tilde{\Omega}_{\text{d}}=0$),
the atom equation (\ref{eq:atom-ME}) gives $\partial_{t}\langle\hat{\tau}_{\mathsf{eb}}^{-}\rangle=-\tilde{\Upsilon}_{\mathsf{eb}}\langle\hat{\tau}_{\mathsf{eb}}^{-}\rangle$,
and $\partial_{t}\langle\hat{\sigma}^{-}\rangle=-\tilde{\Upsilon}_{\mathsf{ab}}\langle\hat{\sigma}^{-}\rangle$
(see also the derivations in Appendix B).}. $\langle\hat{\text{\textsc{n}}}_{\mathsf{a(b)}}\rangle_{\text{ss}}$
is the steady state population on $|\mathsf{a(b)}\rangle$, $\langle\hat{\tau}_{\mathsf{ea}}^{+}\rangle_{\text{ss}}$
is the coherence term, which can be solved by the atom equation (\ref{eq:atom-ME}),
and they give 
\begin{align}
\frac{\langle\hat{\text{\textsc{n}}}_{\mathsf{a}}\rangle_{\text{ss}}}{\langle\hat{\text{\textsc{n}}}_{\mathsf{b}}\rangle_{\text{ss}}} & =\frac{\Gamma_{\mathsf{eb}}^{+}}{\Gamma_{\mathsf{eb}}^{-}}\cdot\frac{\Gamma_{\mathsf{ea}}^{-}+\tilde{\Omega}_{\text{d}}^{2}/2\tilde{\Upsilon}_{\mathsf{ea}}}{\Gamma_{\mathsf{ea}}^{+}+\tilde{\Omega}_{\text{d}}^{2}/2\tilde{\Upsilon}_{\mathsf{ea}}},\quad\frac{\langle\hat{\text{\textsc{n}}}_{\mathsf{e}}\rangle_{\text{ss}}}{\langle\hat{\text{\textsc{n}}}_{\mathsf{b}}\rangle_{\text{ss}}}=e^{-\Omega_{\mathsf{eb}}/T},\nonumber \\
\langle\hat{\tau}_{\mathsf{ea}}^{-}\rangle_{\text{ss}} & =\frac{i\tilde{\Omega}_{\text{d}}}{2\tilde{\Upsilon}_{\mathsf{ea}}}\big(\langle\hat{\text{\textsc{n}}}_{\mathsf{e}}\rangle-\langle\hat{\text{\textsc{n}}}_{\mathsf{a}}\rangle\big).
\end{align}

When the driving strength $\tilde{\Omega}_{\text{d}}$ is strong enough,
the above ratios give $\langle\hat{\text{\textsc{n}}}_{\mathsf{a}}\rangle_{\text{ss}}/\langle\hat{\text{\textsc{n}}}_{\mathsf{b}}\rangle_{\text{ss}}\rightarrow e^{-\Omega_{\mathsf{eb}}/T}\simeq0$
(for the optical frequency $\Omega_{\mathsf{eb}}\gg T\simeq300\,\text{K}$),
namely, $\langle\hat{\text{\textsc{n}}}_{\mathsf{b}}\rangle_{\text{ss}}\simeq1$
and $\langle\hat{\text{\textsc{n}}}_{\mathsf{a,e}}\rangle_{\text{ss}}\simeq\langle\hat{\tau}_{\mathsf{ea}}^{-}\rangle_{\text{ss}}\simeq0$
(see details in Appendix B). That means, the population are fully
concentrated in the ground state $|\mathsf{b}\rangle$. As mentioned
above, such a population distribution effectively gives a zero temperature
$T_{\mathsf{ab}}^{\text{eff}}\rightarrow0$. 

But it is also worth noting that the driving strength $\tilde{\Omega}_{\text{d}}$
also appears in the correction factor $2g^{2}/(\tilde{\Upsilon}_{\mathsf{ab}}+\tilde{\Omega}_{\text{d}}^{2}/4\tilde{\Upsilon}_{\mathsf{eb}})$
in the cooling/heating rates $A_{\pm}$ {[}Eq.\,(\ref{eq:A+-3}){]}.
As a result, when the driving strength $\tilde{\Omega}_{\text{d}}$
is too large, both the cooling and heating rates decrease towards
zero $A_{\pm}\rightarrow0$, and that weakens the cooling effect. 

The reason can be understood by the following picture. For the two
levels $|\mathsf{e}\rangle$ and $|\mathsf{a}\rangle$ under the laser
driving, effectively the driving field also perturbs these two levels,
which makes them shift upwards and downwards. As a result, the energy
gap of the lowest two levels would also be changed correspondingly,
and that makes the energy gap $\Omega_{\mathsf{ab}}$ no longer exactly
resonant with the resonator frequency $\omega_{\text{\textsc{r}}}$.
Because of such an off-resonant coupling, the energy cannot be efficiently
transported from the MW resonator to the atom refrigerator, and thus
the cooling and heating rates are both weakened. 

\begin{figure}
\begin{centering}
\includegraphics[width=0.95\columnwidth]{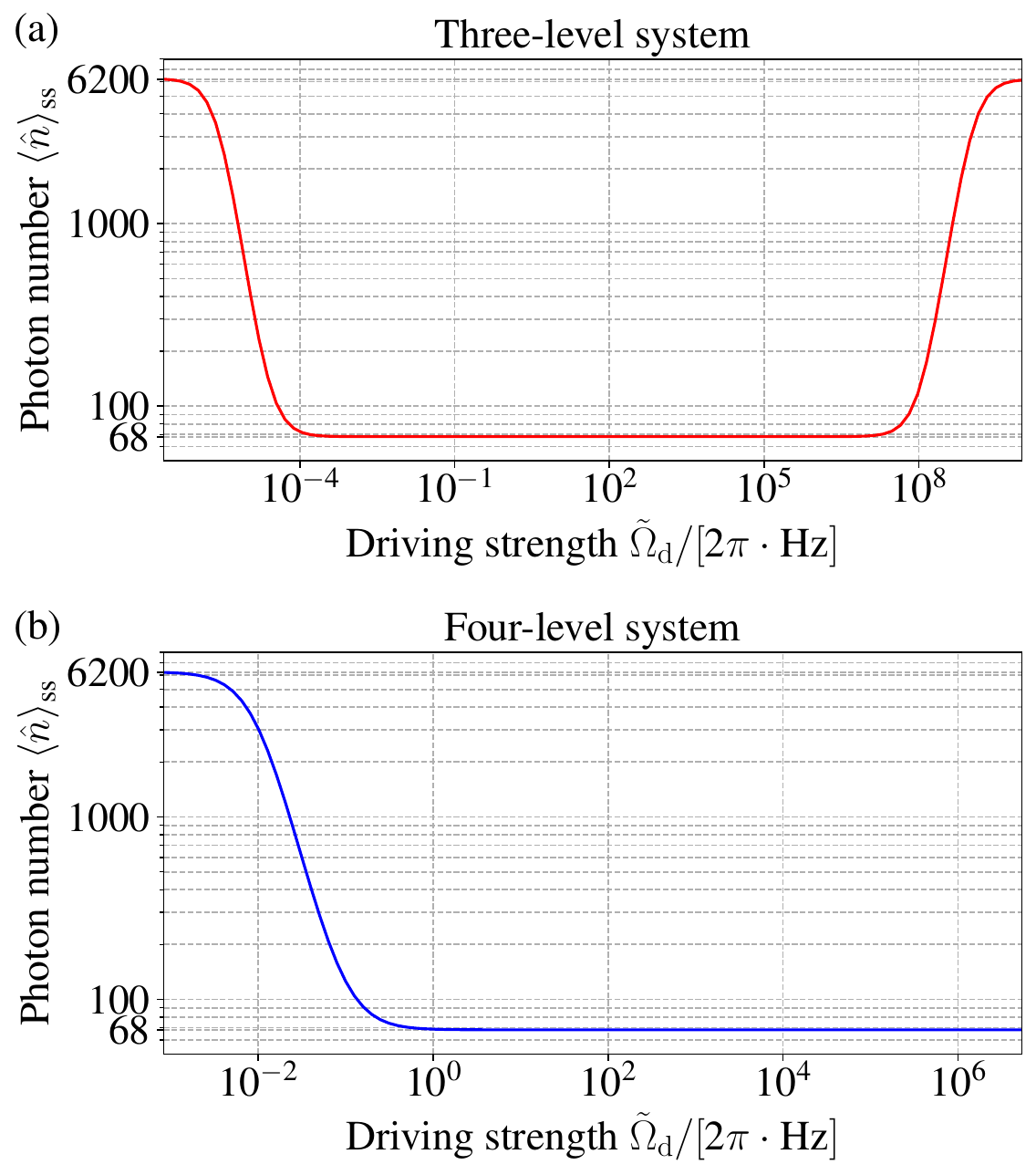}
\par\end{centering}
\caption{The steady photon number $\langle\hat{n}\rangle_{\text{ss}}$ {[}Eq.\,(\ref{eq:nss}){]}
changing with the driving strength $\tilde{\Omega}_{\text{d}}$ for
the (a) three-level and (b) four-level system. The parameters are
set as: resonator frequency $\omega_{\text{\textsc{r}}}/2\pi=1\,\text{GHz}$,
atom-resonator coupling strength $g_{N}/2\pi=1.5\,\text{MHz}$, total
dephasing rate for $|\mathsf{a}\rangle\leftrightarrow|\mathsf{b}\rangle$
$\tilde{\Upsilon}_{\mathsf{ab}}=1\,\text{MHz}$, spontaneous decay
rate $\gamma_{\alpha\beta}\equiv\gamma=1\,\text{GHz}$ (for all the
optical transitions), initial photon number $\bar{\mathsf{n}}_{\text{\textsc{r}}}=6200$
(corresponding to temperature $T\simeq300\,\text{K}$). The frequencies
of the optical transitions are set as $\Omega_{\mathsf{ea}}/2\pi=400\,\text{THz}$
(for the three-level system), $\Omega_{\mathsf{em}}/2\pi=300\,\text{THz}$
and $\Omega_{\mathsf{ma}}/2\pi=400\,\text{THz}$ (for the four-level
system). The numerical minimums in these two cases are both $\text{min}\langle\hat{n}\rangle_{\text{ss}}\simeq68.1$.}

\label{fig-photon}
\end{figure}

In Fig.\,\ref{fig-photon}(a), we show the steady state photon number
$\langle\hat{n}\rangle_{\text{ss}}$ in the MW resonator changing
with the driving strength $\tilde{\Omega}_{\text{d}}$. With the increase
of the driving strength $\tilde{\Omega}_{\text{d}}$, the photon number
$\langle\hat{n}\rangle_{\text{ss}}$ first decreases rapidly, and
finally increases again, which roots from the correction factor in
the cooling rate (\ref{eq:A+-3}) discussed above. Thus, the driving
strength $\tilde{\Omega}_{\text{d}}$ should be controlled within
a finite working region so as to reach the optimum cooling effect.

Now we make an estimation of the cooling limit of $\langle\hat{n}\rangle_{\text{ss}}$
within the working region. From Fig.\,\ref{fig-photon}(a), we can
see indeed a small driving strength $\tilde{\Omega}_{\text{d}}$ is
enough to cool down the atoms to their ground states, which gives
$\langle\hat{\text{\textsc{n}}}_{\mathsf{a}}\rangle_{\text{ss}}\simeq\langle\hat{\tau}_{\mathsf{ea}}^{+}\rangle_{\text{ss}}\simeq0$,
and $\langle\hat{\text{\textsc{n}}}_{\mathsf{b}}\rangle_{\text{ss}}\simeq1$;
meanwhile, such a driving strength is still not large enough to bring
in significant change to the correction factor $2g^{2}/(\tilde{\Upsilon}_{\mathsf{ab}}+\tilde{\Omega}_{\text{d}}^{2}/4\tilde{\Upsilon}_{\mathsf{eb}})$
in $A_{\pm}$, thus that gives $A_{-}\simeq2g^{2}/\tilde{\Upsilon}_{\mathsf{ab}}$,
$A_{+}\simeq0$. Therefore, in the working region, the cooling limit
of $\langle\hat{n}\rangle_{\text{ss}}$ is estimated as {[}from Eq.\,(\ref{eq:nss}){]}
\begin{equation}
\text{min}\langle\hat{n}\rangle_{\text{ss}}\simeq\frac{\kappa\bar{\mathsf{n}}_{\text{\textsc{r}}}}{2g^{2}/\tilde{\Upsilon}_{\mathsf{ab}}+\kappa}\simeq\bar{\mathsf{n}}_{\text{\textsc{r}}}\big/\frac{2g^{2}}{\kappa\tilde{\Upsilon}_{\mathsf{ab}}}.\label{eq:limit-3}
\end{equation}

The upper constraint for the working region of $\tilde{\Omega}_{\text{d}}$
can be roughly estimated from the denominator form of the correction
factor {[}Eq.\,(\ref{eq:A+-3}){]}, which gives significant decrease
when $\tilde{\Upsilon}_{\mathsf{ab}}\simeq\tilde{\Omega}_{\text{d}}^{2}/4\tilde{\Upsilon}_{\mathsf{eb}}$,
and that gives $\tilde{\Omega}_{\text{d}}\simeq2\sqrt{\tilde{\Upsilon}_{\mathsf{eb}}\tilde{\Upsilon}_{\mathsf{ab}}}$.
It turns out such a upper constraint estimation well fits the numerical
result {[}see the position of the increasing part in Fig.\,\ref{fig-photon}(a){]}. 

\section{Four-level refrigerator}

Now we further consider using a four-level system as the refrigerator,
and adopting an indirect pumping approach \citep{wang_four-level_2015,li_quantum_2017,yan_external-level_2021}.
In this setup {[}Fig.\,\ref{fig-demo}(b){]}, still the lowest two
levels $|\mathsf{a,b}\rangle$ are resonantly coupled with the MW
resonator, while a mediated level $|\mathsf{m}\rangle$ is added between
$|\mathsf{e}\rangle$ and $|\mathsf{a}\rangle$ (assuming $\Omega_{\mathsf{em}},\,\Omega_{\mathsf{ma}}$
are in the optical frequency regime), and now the driving laser is
applied to the transition $|\mathsf{e}\rangle\leftrightarrow|\mathsf{m}\rangle$,
which would no longer perturb $|\mathsf{a,b}\rangle$ directly. Similarly
as the above three-level case, only the optical transitions $|\mathsf{e}\rangle\leftrightarrow|\mathsf{m}\rangle$,
$|\mathsf{m}\rangle\leftrightarrow|\mathsf{a}\rangle$, and $|\mathsf{e}\rangle\leftrightarrow|\mathsf{b}\rangle$
are considered {[}the dashed paths in Fig.\,\ref{fig-demo}(b){]},
while the transition between $|\mathsf{a}\rangle$ and $|\mathsf{b}\rangle$
is neglected. The dephasing effects of these levels are also taken
into account.

In this case, the driving laser moves the population from $|\mathsf{m}\rangle$
to $|\mathsf{e}\rangle$, and the ``heat'' could be dumped out through
the optical emission $|\mathsf{e}\rangle\rightarrow|\mathsf{b}\rangle$;
meanwhile, the population decrease in $|\mathsf{m}\rangle$ would
be complemented from $|\mathsf{a}\rangle$ through the thermal excitation
$|\mathsf{m}\rangle\leftrightarrow|\mathsf{a}\rangle$, until their
populations satisfy the Boltzmann distribution. This is similar as
a thermally ``siphonic'' effect. As a result, under a strong enough
driving intensity, the atom population could be fully concentrated
into the ground state $|\mathsf{b}\rangle$, meanwhile, the energy
gap $\Omega_{\mathsf{ab}}$ would no longer be perturbed and still
keep resonant with the MW resonator ($\Omega_{\mathsf{ab}}=\omega_{\text{\textsc{r}}}$). 

Intuitively, the population in $|\mathsf{m}\rangle$ is almost zero,
which makes the driving laser seem being applied on ``nothing''.
But in a finite temperature $T$, there still remains a nonzero population,
though quite small. It turns out this is enough to achieve the above
``siphonic'' cooling process.

For this four-level system, we still apply the adiabatic elimination
to derive the equation for the MW mode alone, and it turns out to
have the same form as the above resonator equation (\ref{eq:mu-t})
for the three-level case. And the heating and cooling rates $A_{\pm}'$
are still defined by Eq.\,(\ref{eq:A}), except now the correlation
functions $\langle\hat{\sigma}^{\pm}(s)\hat{\sigma}^{\mp}(0)\rangle_{\text{ss}}$
in $A_{\pm}'$ {[}Eq.\,(\ref{eq:A}){]} should be calculated from
the four-level system under the laser driving. With the help of the
quantum regression theorem, for this four-level setup, the heating
and cooling rates are obtained as (the full results are presented
in Appendix C) 
\begin{equation}
A_{+}'=\frac{2g^{2}}{\tilde{\Upsilon}_{\mathsf{ab}}'}\langle\hat{\text{\textsc{n}}}_{\mathsf{a}}\rangle_{\text{ss}},\qquad A_{-}'=\frac{2g^{2}}{\tilde{\Upsilon}_{\mathsf{ab}}'}\langle\hat{\text{\textsc{n}}}_{\mathsf{b}}\rangle_{\text{ss}}.\label{eq:A+-4L}
\end{equation}
Here, $\tilde{\Upsilon}_{\mathsf{ab}}':=\frac{1}{2}(\Gamma_{\mathsf{eb}}^{+}+\Gamma_{\mathsf{ma}}^{+}+\tilde{\gamma}_{\text{p}}^{\mathsf{a}}+\tilde{\gamma}_{\text{p}}^{\mathsf{b}})$
is the total dephasing rate for the oscillation between the levels
$|\mathsf{a}\rangle\leftrightarrow|\mathsf{b}\rangle$ (see from the
atom equation $\partial_{t}\langle\hat{\sigma}^{-}\rangle$, similarly
as the above three level case). In the steady state, the population
ratios in this four-level system satisfy (see Appendix C)
\begin{align}
\frac{\langle\hat{\text{\textsc{n}}}_{\mathsf{m}}\rangle_{\text{ss}}}{\langle\hat{\text{\textsc{n}}}_{\mathsf{a}}\rangle_{\text{ss}}} & =e^{-\Omega_{\mathsf{ma}}/T},\hspace*{1em}\frac{\langle\hat{\text{\textsc{n}}}_{\mathsf{e}}\rangle_{\text{ss}}}{\langle\hat{\text{\textsc{n}}}_{\mathsf{b}}\rangle_{\text{ss}}}=e^{-\Omega_{\mathsf{eb}}/T},\nonumber \\
\frac{\langle\hat{\text{\textsc{n}}}_{\mathsf{e}}\rangle_{\text{ss}}}{\langle\hat{\text{\textsc{n}}}_{\mathsf{m}}\rangle_{\text{ss}}} & =\frac{\Gamma_{\mathsf{em}}^{+}+\tilde{\Omega}_{\text{d}}^{2}/2\tilde{\Upsilon}_{\mathsf{ab}}'}{\Gamma_{\mathsf{em}}^{-}+\tilde{\Omega}_{\text{d}}^{2}/2\tilde{\Upsilon}_{\mathsf{ab}}'}.\label{eq:ratio-4}
\end{align}

When the driving strength $\tilde{\Omega}_{\text{d}}\rightarrow\infty$,
the ratios (\ref{eq:ratio-4}) give $\langle\hat{\text{\textsc{n}}}_{\mathsf{e}}\rangle_{\text{ss}}/\langle\hat{\text{\textsc{n}}}_{\mathsf{m}}\rangle_{\text{ss}}\simeq1$,
and $\langle\hat{\text{\textsc{n}}}_{\mathsf{a}}\rangle_{\text{ss}}/\langle\hat{\text{\textsc{n}}}_{\mathsf{b}}\rangle_{\text{ss}}\simeq e^{-(\Omega_{\mathsf{eb}}-\Omega_{\mathsf{ma}})/T}\rightarrow0$.
That indicates the populations could be fully concentrated into the
ground state, i.e., $\langle\hat{\text{\textsc{n}}}_{\mathsf{b}}\rangle_{\text{ss}}\rightarrow1$,
$\langle\hat{\text{\textsc{n}}}_{\mathsf{a}}\rangle_{\text{ss}}\rightarrow0$,
which also gives $T_{\mathsf{ab}}^{\text{eff}}\rightarrow0$.

Unlike the above three-level case {[}Eq.\,(\ref{eq:A+-3}){]}, now
the driving strength $\tilde{\Omega}_{\text{d}}$ no longer appears
in the correction factor $2g^{2}/\tilde{\Upsilon}_{\mathsf{ab}}'$
in $A_{\pm}'$ {[}Eq.\,(\ref{eq:A+-4L}){]}. Thus, with the increase
of the driving light intensity, here the cooling (heating) rate increases
(decreases) monotonically. Therefore, when the driving strength $\tilde{\Omega}_{\text{d}}\rightarrow\infty$,
the cooling performance could achieve the optimum, and that gives
$A_{+}'\rightarrow0$ and $A_{-}'\rightarrow2g^{2}/\tilde{\Upsilon}_{\mathsf{ab}}'$.
In Fig.\,\ref{fig-photon}(b), we see the steady photon number $\langle\hat{n}\rangle_{\text{ss}}$
in the MW resonator decreases monotonically with the increase of the
driving strength $\tilde{\Omega}_{\text{d}}$, and now there is no
upper constraint for $\tilde{\Omega}_{\text{d}}$.

Based on the above results, for this four level system, the cooling
limit of the photon number in the MW resonator {[}Eq.\,(\ref{eq:nss}){]}
is obtained as 
\begin{eqnarray}
\langle\hat{n}\rangle_{\text{ss}} & = & \frac{\frac{2g^{2}}{\kappa\tilde{\Upsilon}_{\mathsf{ab}}'}\langle\hat{\text{\textsc{n}}}_{\mathsf{a}}\rangle+\bar{\mathsf{n}}_{\text{\textsc{r}}}}{\frac{2g^{2}}{\kappa\tilde{\Upsilon}_{\mathsf{ab}}'}\langle\hat{\text{\textsc{n}}}_{\mathsf{b}}-\hat{\text{\textsc{n}}}_{\mathsf{a}}\rangle+1}\nonumber \\
 & \stackrel{\tilde{\Omega}_{\text{d}}\rightarrow\infty}{\longrightarrow} & \frac{\kappa\bar{\mathsf{n}}_{\text{\textsc{r}}}}{2g^{2}/\tilde{\Upsilon}_{\mathsf{ab}}'+\kappa}\simeq\bar{\mathsf{n}}_{\text{\textsc{r}}}\big/\frac{2g^{2}}{\kappa\tilde{\Upsilon}_{\mathsf{ab}}'},\label{eq:limit-4}
\end{eqnarray}
which just has the same form as the above three level case {[}Eq.\,(\ref{eq:limit-3}){]}. 

Here, $\tilde{\Upsilon}_{\mathsf{ab}}'$ is the total dephasing rate
for the oscillation between the MW levels $|\mathsf{a}\rangle\leftrightarrow|\mathsf{b}\rangle$,
which can be measured from the practical decay time of the atom coherence.
From its definition under Eq.\,(\ref{eq:A+-4L}), $\Gamma_{\mathsf{eb}}^{+}\equiv\gamma_{\mathsf{eb}}\,\bar{\mathsf{n}}_{\mathsf{eb}}\simeq0$,
$\Gamma_{\mathsf{ma}}^{+}\equiv\gamma_{\mathsf{ma}}\,\bar{\mathsf{n}}_{\mathsf{ma}}\simeq0$
(since $\Omega_{\mathsf{eb}},\Omega_{\mathsf{ma}}\gg T$), we obtain
$\tilde{\Upsilon}_{\mathsf{ab}}'\simeq\frac{1}{2}(\tilde{\gamma}_{\text{p}}^{\mathsf{a}}+\tilde{\gamma}_{\text{p}}^{\mathsf{b}})$.

All the above discussions are based on the interaction between the
MW resonator and a single multi-level system. Generally, the coupling
strength $g$ between an MW resonator mode and a single atom is quite
small. This can be improved by adopting $N$ atom refrigerators to
couple with the resonator. Correspondingly, the above cooling and
heating rates $A_{\pm}$ obtained from a single refrigerator can be
enlarged by $N$ times. Effectively, this also can be regarded as
the enlargement in the coupling strength, $g\mapsto g_{N}\equiv\sqrt{N}\,g$,
which is similar as the treatment in lasing problems \citep{scully_quantum_1997,breuer_theory_2002,agarwal_quantum_2012}.
Therefore, to achieve a better cooling effect, we need $Ng^{2}/\kappa\tilde{\Upsilon}_{\mathsf{ab}}'\gg1$. 

\section{Experiment estimation}

Now we make an estimation for the possible cooling limit {[}Eq.\,(\ref{eq:limit-4}){]}
in realistic experiments. For the example of an MW resonator with
$\omega_{\text{\textsc{r}}}/2\pi=1\,\text{GHz}$, at room temperature
$T=300\,\text{K}$, the thermal photon number from the surrounding
reservoir is $\bar{\mathsf{n}}_{\text{\textsc{r}}}\simeq6.2\times10^{3}$.
A quality factor $Q\simeq10^{4}$ is an achievable estimation, which
gives the resonator decay rate as $\kappa\simeq0.1\,\text{MHz}$ \citep{breeze_room-temperature_2017,wu_bench-top_2021,day_room-temperature_2024}. 

The multi-level systems can be implemented by the defect structures
in solid crystals (e.g., NV or SiV centers in diamonds or silicon
carbide \citep{fischer_highly_2018,gottscholl_room-temperature_2023,castelletto_quantum_2024},
pentacene molecules doped in the \emph{p}-terphenyl crystal \citep{wu_bench-top_2021,wu_enhanced_2022}).
It is reported that the effective coupling strength between the MW
resonator and NV ensemble could achieve $g_{N}/2\pi\simeq1.5\,\text{MHz}$
\citep{day_room-temperature_2024}. The typical dephasing time of
the MW levels is around $T_{2}^{*}\simeq2\,\mu\text{s}$ \citep{liu_controllable_2012,zhang_arbitrary_2024,breeze_room-temperature_2017},
and that gives the total dephasing rate as $\tilde{\Upsilon}_{\mathsf{ab}}\simeq0.5\,\text{MHz}$. 

Based on the above experimental parameters, the cooling limit (\ref{eq:limit-4})
gives the steady photon number in the MW resonator as $\langle\hat{n}\rangle_{\text{ss}}\simeq68.9$,
which well fits the numerical results in Fig.\,\ref{fig-photon}.
That corresponds to an effective temperature $T_{\text{\textsc{r}}}^{\text{eff}}\equiv\hbar\omega_{\text{\textsc{r}}}\big/k_{\text{\textsc{b}}}\ln(1+1/\langle\hat{n}\rangle_{\text{ss}})\simeq3.3\,\text{K}$
(starting from room temperature). Such a cooling effect is well comparable
with the liquid helium temperature. 

Alternatively, another more specific platform is to use atom gas as
the multilevel systems. For the instance of $^{23}\mathrm{Na}$ atom,
the interaction with the nuclear makes the ground state $3\text{S}_{1/2}$
splitted into two hyperfine levels, whose energy gap is $\Delta\nu\simeq1.77\,\text{GHz}$.
Both these two states can be excited to the state $3\text{P}_{1/2}$
via electric dipole transition, thus such a structure well satisfies
the above three level refrigerator. 

The lowest two atom levels can be coupled with the MW resonator via
the magnetic dipole transition interaction. The transition magnetic
dipole moment of these two hyperfine levels is mainly contributed
from the electron spin, which is similar as the situation of the NV
center. Thus, the transition magnetic dipole moment of these two atom
levels has the similar magnitude as the above NV center. Therefore,
the effective coupling strength between an ensemble of $^{23}\mathrm{Na}$
atoms and the MW resonator could achieve at least the similar strength
as the NV ensemble \citep{day_room-temperature_2024}. Moreover, indeed
generally the volume of the atom gas is much larger than the diamond
crystal, thus the total number $N$ of the atoms could be much larger
than that of NV centers. 

Further, it is worth noting that the cooling limit (\ref{eq:limit-3})
does not depend on the dephasing rate of the two optical transitions,
but only on that of the MW transition. For dilute atom gas, such a
dephasing rate $\tilde{\Upsilon}_{\mathsf{ab}}$ is mainly contributed
from the Doppler broadening. For $^{23}\mathrm{Na}$ atom gas at room
temperature ($\sim300\,\text{K}$), the Doppler broadening for the
MW transition ($\Delta\nu\simeq1.77\,\text{GHz}$) between these two
hyperfine levels, the Doppler broadening\footnote{The Doppler broadening
here is estimated by $\Delta\nu_{\text{D}}\simeq(\nu_{0}/c)\sqrt{8\ln2\,k_{\text{\textsc{b}}}T/m}$,
where the central frequency is taken as $\nu_{0}=1.77\,\text{GHz}$.
} is just $\sim4.6\,\text{kHz}$, which gives a much smaller dephasing
rate $\tilde{\Upsilon}_{\mathsf{ab}}$ than the above solid defect
system. Based on these conditions, the photon number in MW resonator
could be cooled down even below the single photon level, without resorting
to traditional cryogenic systems. 

\begin{figure}
\begin{centering}
\includegraphics[width=0.85\columnwidth]{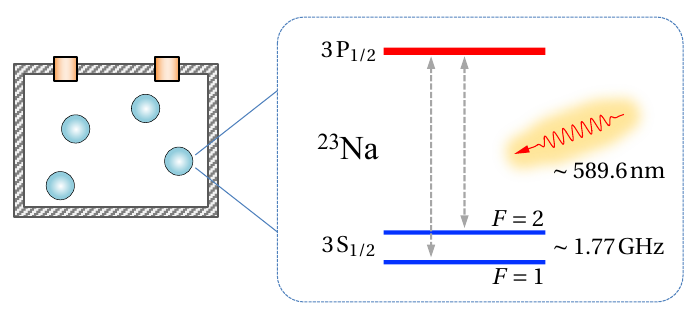}
\par\end{centering}
\caption{Demonstration for atom gas in MW resonator. For the instance of $^{23}\mathrm{Na}$
atom, the interaction with the nuclear makes the ground state $3\text{S}_{1/2}$
splitted into two hyperfine levels ($\Delta\nu\simeq1.77\,\text{GHz}$).
Together with the optical transitions to the excited state, such a
system well satisfies the structure of our three level refrigerator.}

\label{fig-gas}
\end{figure}

\section{Discussions}

In this paper, we consider using three-level or four-level atoms as
quantum refrigerators to cool down an MW resonator, and investigate
the possible cooling limits. Under proper transition structures, a
laser pump drives the atom to work as an SSDG quantum refrigerator,
and the atom population are fully concentrated into the ground state,
effectively giving a zero temperature. Then the thermal photons in
the MW resonator can be continuously absorbed away by the atom. 

By adopting the adiabatic elimination, we obtain a master equation
for the resonator mode, which gives a more precise description for
this cooling system. We find that, for three level systems, though
the atom can be cooled down quite efficiently, a too strong driving
strength also brings in significant perturbation to the atom levels,
which causes the atom-resonator interaction to become off-resonant,
and that weakens the cooling effect. By utilizing an indirect pumping
approach in a four-level refrigerator, such a constraint condition
can be released. We obtain analytical results for the cooling limits
in both cases, which share the same form. Based on some practical
parameters, our estimation shows the cooling limit of the MW resonator
could reach the liquid helium temperature. Our results highlight the
potential of quantum refrigerators as practical, high-performance
solutions for suppressing thermal noise in MW devices without cryogenic
complexity \citep{scully_extracting_2003,quan_quantum_2005,quan_quantum-classical_2006,kosloff_quantum_2013,uzdin_equivalence_2015}.

\vspace{0.5em}\noindent \emph{Acknowledgments }- SWL appreciates
quite much for the helpful discussion with H. Wu, B. Zhang, and D.
Xu in BIT. This study is supported by NSF of China (Grant No. 12475030).

\appendix
\begin{widetext}

\section{The master equation for the MW resonator }

Here we show the derivation of the master equation which describes
the resonator mode alone. In the interaction picture (applied by $\hat{H}_{\text{\textsc{a}}}+\hat{H}_{\text{\textsc{r}}}$),
we rewrite the equation for the atom-resonator system as $\partial_{t}\rho=(\mathcal{L}_{\text{\textsc{a}}}+\mathcal{K}_{\text{\textsc{ar}}}+\mathcal{D}_{\text{\textsc{r}}})[\rho]$,
where 
\begin{align}
\mathcal{L}_{\text{\textsc{a}}}[\rho] & =i[\rho,\,\tilde{V}_{\text{d}}]+\mathcal{D}_{\text{\textsc{a}}}[\rho]+\mathcal{D}_{\text{dep}}[\rho],\nonumber \\
\mathcal{K}_{\text{\textsc{ar}}}[\rho] & =i[\rho,\,\tilde{V}_{\text{\textsc{ar}}}]=i\,[\rho,\,g(\hat{\sigma}^{+}\hat{a}+\hat{\sigma}^{-}\hat{a}^{\dag})].
\end{align}
 Here $\mathcal{D}_{\text{\textsc{r}}}[\rho]$ describes the dissipation
of the resonator mode, $\mathcal{L}_{\text{\textsc{a}}}[\rho]$ describes
the atom dissipation together with the laser driving $\tilde{V}_{\text{d}}$,
and $\mathcal{K}_{\text{\textsc{ar}}}[\rho]$ describes the interaction
between the atom and the resonator.

Here we need a dynamical equation solely for the resonator state $\varrho_{\text{\textsc{r}}}=\mathrm{tr}_{\text{\textsc{a}}}\rho$.
Generally, the dissipation rate of the MW resonator is much slower
than that of the atom, and $\mathcal{D}_{\text{\textsc{r}}}[\rho]$
simply gives $\mathrm{tr}_{\text{\textsc{a}}}\big\{\mathcal{D}_{\text{\textsc{r}}}[\rho]\big\}=\mathcal{D}_{\text{\textsc{r}}}[\varrho_{\text{\textsc{r}}}]$,
therefore, in the following discussions we first omit this term and
then take it back in the final step. 

The master equation $\partial_{t}\rho=(\mathcal{L}_{\text{\textsc{a}}}+\mathcal{K}_{\text{\textsc{ar}}})[\rho]$
has a similar linear structure to the Schr\"odinger equation, thus
we treat $\mathcal{K}_{\text{\textsc{ar}}}$ as a perturbation based
on $\mathcal{L}_{\text{\textsc{a}}}$. The steady states of $\mathcal{L}_{\text{\textsc{a}}}[\rho]=0$
form a degenerated subspace, i.e., $\mathcal{L}_{\text{\textsc{a}}}\left[\ |n\rangle\langle n|\otimes\varrho_{\text{\textsc{a}}}^{\text{ss}}\ \right]=0$,
where $|n\rangle$ are the Fock states for the MW mode, and $\varrho_{\text{\textsc{a}}}^{\text{ss}}$
is the steady state of the atom ($\mathcal{L}_{\text{\textsc{a}}}[\varrho_{\text{\textsc{a}}}^{\text{ss}}]=0$).
Thus, the degenerated perturbation can be applied based on this subspace
\citep{takahashi_half-filled_1977,wu_exact_2022}. Here we introduce
some projection operators $\mathcal{P}[\rho]:=\mathcal{P}_{\text{\textsc{r}}}\cdot\mathcal{P}_{\text{\textsc{a}}}[\rho]$,
where 
\begin{align}
\mathcal{P}_{\text{\textsc{r}}}[\rho] & :=\sum_{n=0}^{\infty}\langle n|\rho|n\rangle\ |n\rangle\langle n|,\qquad\mathcal{P}_{\text{\textsc{a}}}[\rho]:=\lim_{t\rightarrow\infty}e^{t\mathcal{L}_{\text{\textsc{a}}}}[\rho]=\varrho_{\text{\textsc{a}}}^{\text{ss}}\otimes\mathrm{tr}_{\text{\textsc{a}}}[\rho].
\end{align}
 Such a projection gives $\mathcal{P}[\rho(t)]=\varrho_{\text{\textsc{a}}}^{\text{ss}}\otimes\mu(t)$,
where $\mu(t)\equiv\sum p_{n}(t)\,|n\rangle\langle n|$ is the diagonal
part of the resonator state $\varrho_{\text{\textsc{r}}}(t)$. Then
effectively the above master equation can be described by \citep{takahashi_half-filled_1977,wu_exact_2022,cirac_laser_1992,gardiner_quantum_2004}
\begin{align}
\partial_{t}\mathcal{P}[\rho] & =\mathcal{P}\mathcal{K}_{\text{\textsc{ar}}}(-\mathcal{L}_{\text{\textsc{a}}})^{-1}\mathcal{K}_{\text{\textsc{ar}}}\mathcal{P}\,[\rho],\nonumber \\
\Rightarrow\quad\partial_{t}\mu & =\mathcal{P}_{\text{\textsc{r}}}\ \mathrm{tr}_{\text{\textsc{a}}}\Big\{\mathcal{P}_{\text{\textsc{a}}}\ \mathcal{K}_{\text{\textsc{ar}}}(-\mathcal{L}_{\text{\textsc{a}}})^{-1}\mathcal{K}_{\text{\textsc{ar}}}\mathcal{P}[\rho]\Big\}=\mathcal{P}_{\text{\textsc{r}}}\int_{0}^{\infty}ds\,\mathrm{tr}_{\text{\textsc{a}}}\Big\{\mathcal{K}_{\text{\textsc{ar}}}\,e^{s\mathcal{L}_{\text{\textsc{a}}}}\mathcal{K}_{\text{\textsc{ar}}}\,\mathcal{P}[\rho(t)]\Big\}.\label{eq:dtPrho}
\end{align}
In the last equation, the super operator $\mathcal{L}_{\text{\textsc{a}}}^{-1}$
is formally replaced by its Laplace integral, and the integration
term has been simplified by $\mathrm{tr}_{\text{\textsc{a}}}\big\{\mathcal{P}_{\text{\textsc{a}}}\ \mathcal{L}[\rho]\big\}=\mathrm{tr}_{\text{\textsc{a}}}\big\{\,\varrho_{\text{\textsc{a}}}^{\text{ss}}\otimes\mathrm{tr}_{\text{\textsc{a}}}(\mathcal{L}[\rho])\,\big\}=\mathrm{tr}_{\text{\textsc{a}}}\big\{\mathcal{L}[\rho]\big\}$. 

Further, denoting $\Theta_{s}:=e^{s\mathcal{L}_{\text{\textsc{a}}}}\mathcal{K}_{\text{\textsc{ar}}}\mathcal{P}[\rho]$
for short, the above integration term becomes $\mathrm{tr}_{\text{\textsc{a}}}\big\{\mathcal{K}_{\text{\textsc{ar}}}[\Theta_{s}]\big\}=ig\ [\,\mathrm{tr}_{\text{\textsc{a}}}(\hat{\sigma}^{+}\Theta_{s}),\,\hat{a}\,]+ig\ [\,\mathrm{tr}_{\text{\textsc{a}}}(\hat{\sigma}^{-}\Theta_{s}),\,\hat{a}^{\dag}\,]$
(using the relation $\mathrm{tr}_{\text{\textsc{a}}}\big\{\,[\Theta_{s},\,\hat{X}_{\text{\textsc{a}}}\cdot\hat{Y}_{\text{\textsc{r}}}]\,\big\}=[\,\mathrm{tr}_{\text{\textsc{a}}}(\hat{X}_{\text{\textsc{a}}}\Theta_{s}),\,\hat{Y}_{\text{\textsc{r}}}\,]$).
That further gives 
\begin{align}
\mathrm{tr}_{\text{\textsc{a}}}(\hat{\sigma}^{+}\Theta_{s}) & =\mathrm{tr}_{\text{\textsc{a}}}\big\{\hat{\sigma}^{+}\cdot e^{s\mathcal{L}_{\text{\textsc{a}}}}\ \mathcal{K}_{\text{\textsc{ar}}}[\mathcal{P}\rho]\big\}=ig\ \mathrm{tr}_{\text{\textsc{a}}}\Big\{\hat{\sigma}^{+}(s)\cdot[\varrho_{\text{\textsc{a}}}^{\text{ss}}\cdot\mu_{t},\,\hat{\sigma}^{+}\hat{a}+\hat{\sigma}^{-}\hat{a}^{\dag}]\Big\}\nonumber \\
 & =ig\ \Big\{\langle\hat{\sigma}^{+}(0)\hat{\sigma}^{+}(s)\rangle_{\text{ss}}\:\mu_{t}\hat{a}+\langle\hat{\sigma}^{-}(0)\hat{\sigma}^{+}(s)\rangle_{\text{ss}}\:\mu_{t}\hat{a}^{\dag}-\langle\hat{\sigma}^{+}(s)\hat{\sigma}^{+}(0)\rangle_{\text{ss}}\:\hat{a}\mu_{t}-\langle\hat{\sigma}^{+}(s)\hat{\sigma}^{-}(0)\rangle_{\text{ss}}\:\hat{a}^{\dag}\mu_{t}\Big\},\\
\mathrm{tr}_{\text{\textsc{a}}}(\hat{\sigma}^{-}\Theta_{s}) & =ig\ \Big\{\langle\hat{\sigma}^{+}(0)\hat{\sigma}^{-}(s)\rangle_{\text{ss}}\:\mu_{t}\hat{a}+\langle\hat{\sigma}^{-}(0)\hat{\sigma}^{-}(s)\rangle_{\text{ss}}\:\mu_{t}\hat{a}^{\dag}-\langle\hat{\sigma}^{-}(s)\hat{\sigma}^{+}(0)\rangle_{\text{ss}}\:\hat{a}\mu_{t}-\langle\hat{\sigma}^{-}(s)\hat{\sigma}^{-}(0)\rangle_{\text{ss}}\:\hat{a}^{\dag}\mu_{t}\Big\}.\nonumber 
\end{align}

Taking these results back to the above Eq.\,(\ref{eq:dtPrho}), we
obtain 
\begin{align}
\partial_{t}\mu= & ig\,\mathcal{P}_{\text{\textsc{r}}}\int_{0}^{\infty}ds\ \Big\{[\mathrm{tr}_{\text{a}}(\hat{\sigma}^{+}\Theta_{s}),\,\hat{a}]+[\mathrm{tr}_{\text{a}}(\hat{\sigma}^{-}\Theta_{s}),\,\hat{a}^{\dag}]\Big\}\nonumber \\
= & -g^{2}\int_{0}^{\infty}ds\ \Big\{\langle\hat{\sigma}^{-}(0)\hat{\sigma}^{+}(s)\rangle_{\text{ss}}\,[\mu_{t}\hat{a}^{\dag},\,\hat{a}]-\langle\hat{\sigma}^{+}(s)\hat{\sigma}^{-}(0)\rangle_{\text{ss}}\:[\hat{a}^{\dag}\mu_{t},\,\hat{a}]\nonumber \\
 & \quad\qquad\qquad+\langle\hat{\sigma}^{+}(0)\hat{\sigma}^{-}(s)\rangle_{\text{ss}}\:[\mu_{t}\hat{a},\,\hat{a}^{\dag}]-\langle\hat{\sigma}^{-}(s)\hat{\sigma}^{+}(0)\rangle_{\text{ss}}\:[\hat{a}\mu_{t},\,\hat{a}^{\dag}]\Big\},\label{eq:dtmu}
\end{align}
where the double annihilation/creation terms are dropped due to the
projection operation $\mathcal{P}_{\text{\textsc{r}}}$. Since $\left(\langle\hat{\sigma}^{\pm}(0)\hat{\sigma}^{\mp}(s)\rangle_{\text{ss}}\right)^{*}=\langle\hat{\sigma}^{\pm}(s)\hat{\sigma}^{\mp}(0)\rangle_{\text{ss}}$,
the above equation also can be written as 
\begin{align}
\partial_{t}\mu & =\frac{1}{2}A_{+}\big([\hat{a}^{\dag}\mu,\hat{a}]+\mathrm{H.c.}\big)+\frac{1}{2}A_{-}\big([\hat{a}\mu,\hat{a}^{\dag}]+\mathrm{H.c.}\big),\nonumber \\
A_{+} & :=2g^{2}\,\mathrm{Re}\int_{0}^{\infty}ds\,\langle\hat{\sigma}^{+}(s)\hat{\sigma}^{-}(0)\rangle_{\text{ss}},\qquad A_{-}:=2g^{2}\,\mathrm{Re}\int_{0}^{\infty}ds\,\langle\hat{\sigma}^{-}(s)\hat{\sigma}^{+}(0)\rangle_{\text{ss}}.\label{eq:A+-}
\end{align}

Now taking back the original dissipation term $\mathcal{D}_{\text{\textsc{r}}}[\rho]$,
we obtain the master equation solely for the MW resonator, i.e.,
\begin{align}
\partial_{t}\mu & =\Gamma_{+}\big(\hat{a}^{\dagger}\mu\hat{a}-\frac{1}{2}\{\mu,\,\hat{a}^{\dagger}\hat{a}\}\big)+\Gamma_{-}\big(\hat{a}\mu\hat{a}^{\dag}-\frac{1}{2}\{\mu,\,\hat{a}\hat{a}^{\dagger}\}\big),\\
\Gamma_{+} & =A_{+}+\kappa\bar{\mathsf{n}}_{\text{\textsc{r}}},\qquad\Gamma_{-}=A_{-}+\kappa(\bar{\mathsf{n}}_{\text{\textsc{r}}}+1).\nonumber 
\end{align}
Here $\Gamma_{+\,(-)}$ indicates the increasing (decreasing) rate
of the photon number of the resonator, and $A_{+\,(-)}$ can be regarded
as the heating (cooling) rate induced by the atom. In the steady state
$t\rightarrow\infty$, that gives the MW photon number as 
\begin{equation}
\langle\hat{n}\rangle_{\text{ss}}=\frac{\Gamma_{+}}{\Gamma_{-}-\Gamma_{+}}=\frac{A_{+}+\kappa\bar{\mathsf{n}}_{\text{\textsc{r}}}}{A_{-}-A_{+}+\kappa}.
\end{equation}
If the cooling rate is fast enough ($A_{-}\gg A_{+},\kappa\bar{\mathsf{n}}_{\text{\textsc{r}}}$),
the MW photon number becomes $\langle\hat{n}\rangle_{\text{ss}}\rightarrow0$,
which means the thermal noise in the resonator can be greatly suppressed. 

The above derivations for the cooling and heating rates are based
on the interaction between the MW mode and a single atom. A large
number of $N$ atoms can be placed in the MW resonator and used to
absorb the thermal photons together. In this case, the cooling and
heating rates can be enlarged by $N$ times ($A_{\pm}\mapsto NA_{\pm}$),
or equivalently, the atom-resonator coupling strength $g$ can be
regarded as enlarged by $\sqrt{N}$ times ($g\mapsto g_{N}:=\sqrt{N}\,g$). 

\section{The three level system under driving}

\subsection{Steady state expectations }

Here we study the behavior of the three level system when there is
no interaction with the resonator. A driving laser is applied to the
transition path $|\mathsf{e}\rangle\leftrightarrow|\mathsf{a}\rangle$,
and the atom dynamics is described by the master equation (interaction
picture) 
\begin{align}
\partial_{t}\varrho_{\text{\textsc{a}}} & =i[\varrho_{\text{\textsc{a}}},\,\frac{1}{2}\tilde{\Omega}_{\text{d}}(\hat{\tau}_{\mathsf{ea}}^{+}+\hat{\tau}_{\mathsf{ea}}^{-})]+\mathcal{D}_{\text{\textsc{a}}}[\varrho_{\text{\textsc{a}}}]+\mathcal{D}_{\text{dep}}[\varrho_{\text{\textsc{a}}}],\label{eq:ME-atom}\\
\mathcal{D}_{\text{\textsc{a}}}[\varrho_{\text{\textsc{a}}}] & =\sum_{\alpha,\beta}^{\varepsilon_{\alpha}>\varepsilon_{\beta}}\Gamma_{\alpha\beta}^{+}\big(\hat{\tau}_{\alpha\beta}^{+}\varrho_{\text{\textsc{a}}}\hat{\tau}_{\alpha\beta}^{-}-\frac{1}{2}\{\hat{\tau}_{\alpha\beta}^{-}\hat{\tau}_{\alpha\beta}^{+},\,\varrho_{\text{\textsc{a}}}\}\big)+\Gamma_{\alpha\beta}^{-}\big(\hat{\tau}_{\alpha\beta}^{-}\varrho_{\text{\textsc{a}}}\hat{\tau}_{\alpha\beta}^{+}-\frac{1}{2}\{\hat{\tau}_{\alpha\beta}^{+}\hat{\tau}_{\alpha\beta}^{-},\,\varrho_{\text{\textsc{a}}}\}\big),\nonumber \\
\mathcal{D}_{\text{dep}}[\varrho_{\text{\textsc{a}}}] & =\sum_{\alpha}\tilde{\gamma}_{\text{p}}^{\alpha}\big(\hat{\text{\textsc{n}}}_{\alpha}\varrho_{\text{\textsc{a}}}\hat{\text{\textsc{n}}}_{\alpha}-\tfrac{1}{2}\{\hat{\text{\textsc{n}}}_{\alpha},\varrho_{\text{\textsc{a}}}\}\big).\nonumber 
\end{align}
The transition structure of $\mathcal{D}_{\text{\textsc{a}}}[\varrho_{\text{\textsc{a}}}]$
is demonstrated in Fig.\,1(a) in the main text (for $|\mathsf{e}\rangle\leftrightarrow|\mathsf{a}\rangle$
and $|\mathsf{e}\rangle\leftrightarrow|\mathsf{b}\rangle$). For the
transition $|\alpha\rangle\leftrightarrow|\beta\rangle$ ($\varepsilon_{\alpha}>\varepsilon_{\beta}$),
we denote $\hat{\tau}_{\alpha\beta}^{+}:=|\alpha\rangle\langle\beta|=(\hat{\tau}_{\alpha\beta}^{-})^{\dag}$
as the transition operators; $\Gamma_{\alpha\beta}^{+}=\gamma_{\alpha\beta}\bar{\mathsf{n}}_{\alpha\beta}$,
$\Gamma_{\alpha\beta}^{-}=\gamma_{\alpha\beta}(\bar{\mathsf{n}}_{\alpha\beta}+1)$
are the dissipation rates, with $\bar{\mathsf{n}}_{\alpha\beta}\equiv(e^{\Omega_{\alpha\beta}/T}-1)^{-1}$
and $\Omega_{\alpha\beta}\equiv\varepsilon_{\alpha}-\varepsilon_{\beta}$,
which satisfy $\Gamma_{\alpha\beta}^{+}/\Gamma_{\alpha\beta}^{-}=e^{-\Omega_{\alpha\beta}/T}$.
The energy gap between $|\mathsf{a}\rangle$ and $|\mathsf{b}\rangle$
is in the MW regime, thus the spontaneous decay rate between these
two levels is generally negligibly small. And $\mathcal{D}_{\text{dep}}[\varrho_{\text{\textsc{a}}}]$
describes the pure dephasing effects of the atom levels, where $\hat{\text{\textsc{n}}}_{\alpha}:=|\alpha\rangle\langle\alpha|$,
and $\tilde{\gamma}_{\text{p}}^{\alpha}$ are the pure dephasing rates.

From the above master equation (\ref{eq:ME-atom}) it turns out the
equations of $\langle\hat{\tau}_{\mathsf{ea}}^{\pm}\rangle$, $\langle\hat{\text{\textsc{n}}}_{\mathsf{e,a,b}}\rangle$
($\hat{\text{\textsc{n}}}_{\alpha}:=|\alpha\rangle\langle\alpha|$)
form a closed set, i.e., 
\begin{align}
\partial_{t}\langle\hat{\tau}_{\mathsf{ea}}^{-}\rangle & =+\frac{i}{2}\tilde{\Omega}_{\text{d}}\big(\langle\hat{\text{\textsc{n}}}_{\mathsf{e}}\rangle-\langle\hat{\text{\textsc{n}}}_{\mathsf{a}}\rangle\big)-\frac{1}{2}(\Gamma_{\mathsf{ea}}^{+}+\Gamma_{\mathsf{ea}}^{-}+\Gamma_{\mathsf{eb}}^{-}+\tilde{\gamma}_{\text{p}}^{\mathsf{e}}+\tilde{\gamma}_{\text{p}}^{\mathsf{a}})\langle\hat{\tau}_{\mathsf{ea}}^{-}\rangle:=\frac{i}{2}\tilde{\Omega}_{\text{d}}\big(\langle\hat{\text{\textsc{n}}}_{\mathsf{e}}\rangle-\langle\hat{\text{\textsc{n}}}_{\mathsf{a}}\rangle\big)-\tilde{\Upsilon}_{\mathsf{ea}}\langle\hat{\tau}_{\mathsf{ea}}^{-}\rangle,\nonumber \\
\partial_{t}\langle\hat{\text{\textsc{n}}}_{\mathsf{e}}\rangle & =\big[\Gamma_{\mathsf{eb}}^{+}\langle\hat{\text{\textsc{n}}}_{\mathsf{b}}\rangle-\Gamma_{\mathsf{eb}}^{-}\langle\hat{\text{\textsc{n}}}_{\mathsf{e}}\rangle\big]+\big[\Gamma_{\mathsf{ea}}^{+}\langle\hat{\text{\textsc{n}}}_{\mathsf{a}}\rangle-\Gamma_{\mathsf{ea}}^{-}\langle\hat{\text{\textsc{n}}}_{\mathsf{e}}\rangle\big]+\frac{i}{2}\tilde{\Omega}_{\text{d}}\big(\langle\hat{\tau}_{\mathsf{ea}}^{-}\rangle-\langle\hat{\tau}_{\mathsf{ea}}^{+}\rangle\big),\label{eq:steady}\\
\partial_{t}\langle\hat{\text{\textsc{n}}}_{\mathsf{b}}\rangle & =-\big[\Gamma_{\mathsf{eb}}^{+}\langle\hat{\text{\textsc{n}}}_{\mathsf{b}}\rangle-\Gamma_{\mathsf{eb}}^{-}\langle\hat{\text{\textsc{n}}}_{\mathsf{e}}\rangle\big].\nonumber 
\end{align}
Here we denote $\tilde{\Upsilon}_{\mathsf{ea}}:=\frac{1}{2}(\Gamma_{\mathsf{ea}}^{+}+\Gamma_{\mathsf{ea}}^{-}+\Gamma_{\mathsf{eb}}^{-}+\tilde{\gamma}_{\text{p}}^{\mathsf{e}}+\tilde{\gamma}_{\text{p}}^{\mathsf{a}})$.
The above equation $\partial_{t}\langle\hat{\tau}_{\mathsf{ea}}^{-}\rangle$
indicates $\tilde{\Upsilon}_{\mathsf{ea}}$ is just the total dephasing
rate for the oscillation between the levels $|\mathsf{e}\rangle$
and $|\mathsf{a}\rangle$, which sums up the contributions from $\Gamma_{\alpha\beta}^{\pm}$
and $\tilde{\gamma}_{\text{p}}^{\alpha}$. In the steady state, the
time derivatives all give zero, and their steady states give 
\begin{equation}
\frac{\langle\hat{\text{\textsc{n}}}_{\mathsf{e}}\rangle_{\text{ss}}}{\langle\hat{\text{\textsc{n}}}_{\mathsf{b}}\rangle_{\text{ss}}}=\frac{\Gamma_{\mathsf{eb}}^{+}}{\Gamma_{\mathsf{eb}}^{-}}=e^{-\Omega_{\mathsf{eb}}/T},\qquad\frac{\langle\hat{\text{\textsc{n}}}_{\mathsf{a}}\rangle_{\text{ss}}}{\langle\hat{\text{\textsc{n}}}_{\mathsf{b}}\rangle_{\text{ss}}}=\frac{\Gamma_{\mathsf{eb}}^{+}}{\Gamma_{\mathsf{eb}}^{-}}\cdot\frac{\Gamma_{\mathsf{ea}}^{-}+\tilde{\Omega}_{\text{d}}^{2}/2\tilde{\Upsilon}_{\mathsf{ea}}}{\Gamma_{\mathsf{ea}}^{+}+\tilde{\Omega}_{\text{d}}^{2}/2\tilde{\Upsilon}_{\mathsf{ea}}},\qquad\langle\hat{\tau}_{\mathsf{ea}}^{-}\rangle_{\text{ss}}=\frac{i\tilde{\Omega}_{\text{d}}}{2\tilde{\Upsilon}_{\mathsf{ea}}}\big(\langle\hat{\text{\textsc{n}}}_{\mathsf{e}}\rangle_{\text{ss}}-\langle\hat{\text{\textsc{n}}}_{\mathsf{a}}\rangle_{\text{ss}}\big).\label{eq:N-3L}
\end{equation}
Together with the condition $\langle\hat{\text{\textsc{n}}}_{\mathsf{e}}\rangle_{\text{ss}}+\langle\hat{\text{\textsc{n}}}_{\mathsf{a}}\rangle_{\text{ss}}+\langle\hat{\text{\textsc{n}}}_{\mathsf{b}}\rangle_{\text{ss}}=1$,
the above steady state values can be well obtained. When there is
no driving light ($\tilde{\Omega}_{\text{d}}\rightarrow0$), the population
ratio $\langle\hat{\text{\textsc{n}}}_{\mathsf{a}}\rangle_{\text{ss}}/\langle\hat{\text{\textsc{n}}}_{\mathsf{b}}\rangle_{\text{ss}}=\Gamma_{\mathsf{eb}}^{+}\Gamma_{\mathsf{ea}}^{-}/\Gamma_{\mathsf{eb}}^{-}\Gamma_{\mathsf{ea}}^{+}=e^{-\omega_{\text{\textsc{r}}}/T}$
well returns the Boltzmann distribution. When the driving strength
is quite strong ($\tilde{\Omega}_{\text{d}}\rightarrow\infty$), this
population ratio becomes $\langle\hat{\text{\textsc{n}}}_{\mathsf{a}}\rangle_{\text{ss}}/\langle\hat{\text{\textsc{n}}}_{\mathsf{b}}\rangle_{\text{ss}}\rightarrow e^{-\Omega_{\mathsf{eb}}/T}$.
That means, under room temperature ($T\simeq300\,\text{K}$), if $\Omega_{\mathsf{eb}}$
is in the optical frequency regime ($\Omega_{\mathsf{eb}}/T\gg1$),
we have $\langle\hat{\text{\textsc{n}}}_{\mathsf{b}}\rangle_{\text{ss}}\simeq1$
and $\langle\hat{\text{\textsc{n}}}_{\mathsf{b,e}}\rangle_{\text{ss}}\simeq\langle\hat{\tau}_{\mathsf{ea}}^{-}\rangle_{\text{ss}}\simeq0$
{[}see Eq.\,(\ref{eq:N-3L}){]}, which means the atom population
is fully concentrated in the ground state $|\mathsf{b}\rangle$. 

\subsection{Time correlation functions }

To calculate the cooling and heating rates $A_{\pm}$ from the correlation
functions {[}Eq.\,(\ref{eq:A+-}){]}, we need to study the equations
of $\langle\hat{\sigma}^{\pm}(t)\rangle$ {[}denoting $\hat{\sigma}^{+}:=|\mathsf{a}\rangle\langle\mathsf{b}|=(\hat{\sigma}^{-})^{\dag}${]},
and that gives 
\begin{align}
\partial_{t}\langle\hat{\sigma}^{+}\rangle & =\frac{i}{2}\tilde{\Omega}_{\text{d}}\langle\hat{\tau}_{\mathsf{eb}}^{+}\rangle-\frac{1}{2}(\Gamma_{\mathsf{ea}}^{+}+\Gamma_{\mathsf{eb}}^{+}+\tilde{\gamma}_{\text{p}}^{\mathsf{a}}+\tilde{\gamma}_{\text{p}}^{\mathsf{b}})\langle\hat{\sigma}^{+}\rangle:=\frac{i}{2}\tilde{\Omega}_{\text{d}}\langle\hat{\tau}_{\mathsf{eb}}^{+}\rangle-\tilde{\Upsilon}_{\mathsf{ab}}\langle\hat{\sigma}^{+}\rangle,\nonumber \\
\partial_{t}\langle\hat{\tau}_{\mathsf{eb}}^{+}\rangle & =\frac{i}{2}\tilde{\Omega}_{\text{d}}\langle\hat{\sigma}^{+}\rangle-\frac{1}{2}(\Gamma_{\mathsf{ea}}^{-}+\Gamma_{\mathsf{eb}}^{-}+\Gamma_{\mathsf{eb}}^{+}+\tilde{\gamma}_{\text{p}}^{\mathsf{e}}+\tilde{\gamma}_{\text{p}}^{\mathsf{b}})\langle\hat{\tau}_{\mathsf{eb}}^{+}\rangle:=\frac{i}{2}\tilde{\Omega}_{\text{d}}\langle\hat{\sigma}^{+}\rangle-\tilde{\Upsilon}_{\mathsf{eb}}\langle\hat{\tau}_{\mathsf{eb}}^{+}\rangle.\label{eq:Teb}
\end{align}
Here we denote $\tilde{\Upsilon}_{\mathsf{ab}}:=\frac{1}{2}(\Gamma_{\mathsf{ea}}^{+}+\Gamma_{\mathsf{eb}}^{+}+\tilde{\gamma}_{\text{p}}^{\mathsf{a}}+\tilde{\gamma}_{\text{p}}^{\mathsf{b}})$
and $\tilde{\Upsilon}_{\mathsf{eb}}:=\frac{1}{2}(\Gamma_{\mathsf{ea}}^{-}+\Gamma_{\mathsf{eb}}^{-}+\Gamma_{\mathsf{eb}}^{+}+\tilde{\gamma}_{\text{p}}^{\mathsf{e}}+\tilde{\gamma}_{\text{p}}^{\mathsf{b}})$.
These two equations indicate that $\tilde{\Upsilon}_{\mathsf{ab}}$
and $\tilde{\Upsilon}_{\mathsf{eb}}$ are just the total dephasing
rates for the oscillations $|\mathsf{a}\rangle\leftrightarrow|\mathsf{b}\rangle$
and $|\mathsf{e}\rangle\leftrightarrow|\mathsf{b}\rangle$ respectively.
Denoting $\mathbf{v}_{t}:=\big(\,\langle\hat{\sigma}_{\mathsf{ab}}^{+}(t)\rangle,\,\langle\hat{\tau}_{\mathsf{eb}}^{+}(t)\rangle\,\big)^{T}$,
these two equations also can be written as 
\begin{equation}
\partial_{t}\mathbf{v}_{t}=\mathbf{G}\cdot\mathbf{v}_{t},\qquad\mathbf{G}:=\left[\begin{array}{cc}
-\tilde{\Upsilon}_{\mathsf{ab}} & i\tilde{\Omega}_{\text{d}}/2\\
i\tilde{\Omega}_{\text{d}}/2 & -\tilde{\Upsilon}_{\mathsf{eb}}
\end{array}\right].
\end{equation}

Then the correlation function $\langle\hat{\sigma}^{+}(t)\hat{\sigma}^{-}(0)\rangle_{\text{ss}}$
can be calculated with the help of the quantum regression theorem
\citep{scully_quantum_1997,agarwal_quantum_2012,breuer_theory_2002}.
Denoting $\mathbf{V}_{t}:=\big(\,\langle\hat{\sigma}^{+}(t)\hat{\sigma}^{-}(0)\rangle_{\text{ss}},\,\langle\hat{\tau}_{\mathsf{eb}}^{+}(t)\hat{\sigma}^{-}(0)\rangle_{\text{ss}}\,\big)^{T}$,
which satisfies $\mathbf{V}_{t\rightarrow\infty}=(0,0)^{T}$, the
quantum regression theorem states that $\mathbf{V}_{t}$ has the same
equation form as that of $\mathbf{v}_{t}$ {[}Eq.\,(\ref{eq:Teb}){]},
i.e., $\partial_{t}\mathbf{V}_{t}=\mathbf{G}\cdot\mathbf{V}_{t}$.
Thus, the correlation function $\langle\hat{\sigma}^{+}(t)\hat{\sigma}^{-}(0)\rangle_{\text{ss}}$
can be obtained as the first component of $\mathbf{V}_{t}=e^{\mathbf{G}t}\cdot\mathbf{V}_{0}$,
where $\mathbf{V}_{0}=\big(\,\langle\hat{\text{\textsc{n}}}_{\mathsf{a}}\rangle_{\text{ss}},\,\langle\hat{\tau}_{\mathsf{ea}}^{+}\rangle_{\text{ss}}\,\big)^{T}$.
Then the time integration of $\langle\hat{\sigma}^{+}(t)\hat{\sigma}^{-}(0)\rangle_{\text{ss}}$
can be directly obtained as the first component of 
\begin{equation}
\int_{0}^{\infty}dt\,\mathbf{V}_{t}=\int_{0}^{\infty}dt\,e^{\mathbf{G}t}\cdot\mathbf{V}_{0}=-\mathbf{G}^{-1}\cdot\mathbf{V}_{0}.\label{eq:Vt=00003DG.V0}
\end{equation}
 As a result, the heating rate {[}Eq.\,(\ref{eq:A+-}){]} is obtained
as 
\begin{align}
A_{+} & =\frac{2g^{2}}{\tilde{\Upsilon}_{\mathsf{ab}}+\tilde{\Omega}_{\text{d}}^{2}/4\tilde{\Upsilon}_{\mathsf{eb}}}\mathrm{Re}\big[\langle\hat{\text{\textsc{n}}}_{\mathsf{a}}\rangle_{\text{ss}}+\frac{i\tilde{\Omega}_{\text{d}}}{2\tilde{\Upsilon}_{\mathsf{eb}}}\langle\hat{\tau}_{\mathsf{ea}}^{+}\rangle_{\text{ss}}\big]\nonumber \\
 & =\frac{2g^{2}}{\tilde{\Upsilon}_{\mathsf{ab}}+\tilde{\Omega}_{\text{d}}^{2}/4\tilde{\Upsilon}_{\mathsf{eb}}}\cdot\frac{\Gamma_{\mathsf{ea}}^{-}+(1-\gamma_{\mathsf{ea}}/2\tilde{\Upsilon}_{\mathsf{eb}})\,\tilde{\Omega}_{\text{d}}^{2}/2\tilde{\Upsilon}_{\mathsf{ea}}}{(1+e^{\frac{\Omega_{\mathsf{ea}}}{T}}+e^{\frac{\Omega_{\mathsf{eb}}}{T}})\Gamma_{\mathsf{ea}}^{+}+(2+e^{\frac{\Omega_{\mathsf{eb}}}{T}})\,\tilde{\Omega}_{\text{d}}^{2}/2\tilde{\Upsilon}_{\mathsf{ea}}}.\label{eq:A+}
\end{align}

Similarly, the correlation function $\langle\hat{\sigma}^{-}(t)\hat{\sigma}^{+}(0)\rangle_{\text{ss}}$
is calculated in the same way, where the above vectors $\mathbf{V}_{t,\,0}$
and matrix $\mathbf{G}$ should be changed to be 
\begin{equation}
\mathbf{V}_{t}:=\big(\,\langle\hat{\sigma}_{\mathsf{ab}}^{-}(t)\hat{\sigma}^{+}(0)\rangle_{\text{ss}},\,\langle\hat{\tau}_{\mathsf{eb}}^{-}(t)\hat{\sigma}^{+}(0)\rangle_{\text{ss}}\,\big)^{T},\qquad\mathbf{V}_{0}=\big(\,\langle\hat{\text{\textsc{n}}}_{\mathsf{b}}\rangle_{\text{ss}},\,0\,\big)^{T},\qquad\mathbf{G}=\left[\begin{array}{cc}
-\tilde{\Upsilon}_{\mathsf{ab}} & -i\tilde{\Omega}_{\text{d}}/2\\
-i\tilde{\Omega}_{\text{d}}/2 & -\tilde{\Upsilon}_{\mathsf{eb}}
\end{array}\right].
\end{equation}
And that gives the cooling rate {[}Eq.\,(\ref{eq:A+-}){]} as 
\begin{align}
A_{-} & =\frac{2g^{2}}{\bar{\Upsilon}_{\mathsf{ab}}+\tilde{\Omega}_{\text{d}}^{2}/4\tilde{\Upsilon}_{\mathsf{eb}}}\langle\hat{\text{\textsc{n}}}_{\mathsf{b}}\rangle_{\text{ss}}\nonumber \\
 & =\frac{2g^{2}}{\tilde{\Upsilon}_{\mathsf{ab}}+\tilde{\Omega}_{\text{d}}^{2}/4\tilde{\Upsilon}_{\mathsf{eb}}}\cdot\frac{e^{\frac{\Omega_{\mathsf{eb}}}{T}}(\Gamma_{\mathsf{ea}}^{+}+\tilde{\Omega}_{\text{d}}^{2}/2\tilde{\Upsilon}_{\mathsf{ea}})}{(1+e^{\frac{\Omega_{\mathsf{ea}}}{T}}+e^{\frac{\Omega_{\mathsf{eb}}}{T}})\Gamma_{\mathsf{ea}}^{+}+(2+e^{\frac{\Omega_{\mathsf{eb}}}{T}})\,\tilde{\Omega}_{\text{d}}^{2}/2\tilde{\Upsilon}_{\mathsf{ea}}}.\label{eq:A-}
\end{align}

When there is no driving light ($\tilde{\Omega}_{\text{d}}\rightarrow0$),
the cooling and heating rates give $A_{-}/A_{+}=\langle\hat{\text{\textsc{n}}}_{\mathsf{b}}\rangle_{\text{ss}}/\langle\hat{\text{\textsc{n}}}_{\mathsf{a}}\rangle_{\text{ss}}=e^{\omega_{\text{\textsc{r}}}/T}$,
which naturally returns to the Boltzmann ratio, and that indicates
the cooling and heating effect to the MW resonator is the same with
the contribution of the surrounding bath with temperature $T$, which
keeps the photon number in the resonator as $\langle\hat{n}\rangle_{\text{ss}}=\bar{\mathsf{n}}_{\text{\textsc{r}}}$.
With the increase of the driving strength, the cooling rate $A_{-}$
firstly increases, but then decreases towards zero due to the correction
factor $2g^{2}/(\tilde{\Upsilon}_{\mathsf{ab}}+\tilde{\Omega}_{\text{d}}^{2}/4\tilde{\Upsilon}_{\mathsf{eb}})$,
and that weakens the cooling effect. 

\section{The four level system under driving}

Here we study the behavior of the four level system when there is
no interaction with the resonator. A driving laser is applied to the
transition path $|\mathsf{e}\rangle\leftrightarrow|\mathsf{m}\rangle$,
and the atom dynamics is described by the master equation (interaction
picture) 
\begin{equation}
\partial_{t}\varrho_{\text{\textsc{a}}}=i[\varrho_{\text{\textsc{a}}},\,\frac{1}{2}\tilde{\Omega}_{\text{d}}(\hat{\tau}_{\mathsf{em}}^{+}+\hat{\tau}_{\mathsf{em}}^{-})]+\mathcal{D}_{\text{\textsc{a}}}[\varrho_{\text{\textsc{a}}}]+\mathcal{D}_{\text{dep}}[\varrho_{\text{\textsc{a}}}].\label{eq:ME-atom-4L}
\end{equation}
$\mathcal{D}_{\text{\textsc{a}}}[\varrho_{\text{\textsc{a}}}]$ describes
the transitions for $|\mathsf{e}\rangle\leftrightarrow|\mathsf{b}\rangle$,
$|\mathsf{e}\rangle\leftrightarrow|\mathsf{m}\rangle$ and $|\mathsf{m}\rangle\leftrightarrow|\mathsf{a}\rangle$
{[}see Fig.\,1(b) in the main text{]}. Then we obtain the equations
of $\langle\hat{\tau}_{\mathsf{em}}^{\pm}\rangle$, $\langle\hat{\text{\textsc{n}}}_{\mathsf{e,m,a,b}}\rangle$,
i.e., 
\begin{align}
\partial_{t}\langle\hat{\tau}_{\mathsf{em}}^{-}\rangle & =+\frac{i}{2}\tilde{\Omega}_{\text{d}}\big(\langle\hat{\text{\textsc{n}}}_{\mathsf{e}}\rangle-\langle\hat{\text{\textsc{n}}}_{\mathsf{m}}\rangle\big)-\frac{1}{2}(\Gamma_{\mathsf{em}}^{+}+\Gamma_{\mathsf{em}}^{-}+\Gamma_{\mathsf{eb}}^{-}+\Gamma_{\mathsf{ma}}^{-}+\tilde{\gamma}_{\text{p}}^{\mathsf{e}}+\tilde{\gamma}_{\text{p}}^{\mathsf{m}})\langle\hat{\tau}_{\mathsf{em}}^{-}\rangle:=\frac{i}{2}\tilde{\Omega}_{\text{d}}\big(\langle\hat{\text{\textsc{n}}}_{\mathsf{e}}\rangle-\langle\hat{\text{\textsc{n}}}_{\mathsf{m}}\rangle\big)-\tilde{\Upsilon}_{\mathsf{em}}'\,\langle\hat{\tau}_{\mathsf{em}}^{-}\rangle,\nonumber \\
\partial_{t}\langle\hat{\text{\textsc{n}}}_{\mathsf{e}}\rangle & =\big[\Gamma_{\mathsf{eb}}^{+}\langle\hat{\text{\textsc{n}}}_{\mathsf{b}}\rangle-\Gamma_{\mathsf{eb}}^{-}\langle\hat{\text{\textsc{n}}}_{\mathsf{e}}\rangle\big]+\big[\Gamma_{\mathsf{em}}^{+}\langle\hat{\text{\textsc{n}}}_{\mathsf{m}}\rangle-\Gamma_{\mathsf{em}}^{-}\langle\hat{\text{\textsc{n}}}_{\mathsf{e}}\rangle\big]+\frac{i}{2}\tilde{\Omega}_{\text{d}}\big(\langle\hat{\tau}_{\mathsf{em}}^{-}\rangle-\langle\hat{\tau}_{\mathsf{em}}^{+}\rangle\big),\nonumber \\
\partial_{t}\langle\hat{\text{\textsc{n}}}_{\mathsf{a}}\rangle & =\Gamma_{\mathsf{ma}}^{-}\langle\hat{\text{\textsc{n}}}_{\mathsf{m}}\rangle-\Gamma_{\mathsf{ma}}^{+}\langle\hat{\text{\textsc{n}}}_{\mathsf{a}}\rangle,\nonumber \\
\partial_{t}\langle\hat{\text{\textsc{n}}}_{\mathsf{b}}\rangle & =\Gamma_{\mathsf{eb}}^{-}\langle\hat{\text{\textsc{n}}}_{\mathsf{e}}\rangle-\Gamma_{\mathsf{eb}}^{+}\langle\hat{\text{\textsc{n}}}_{\mathsf{b}}\rangle.\label{eq:steady-4L}
\end{align}
Here, we denote $\tilde{\Upsilon}_{\mathsf{em}}'=\frac{1}{2}(\Gamma_{\mathsf{em}}^{+}+\Gamma_{\mathsf{em}}^{-}+\Gamma_{\mathsf{eb}}^{-}+\Gamma_{\mathsf{ma}}^{-}+\tilde{\gamma}_{\text{p}}^{\mathsf{e}}+\tilde{\gamma}_{\text{p}}^{\mathsf{m}})$,
and the above equation $\partial_{t}\langle\hat{\tau}_{\mathsf{em}}^{-}\rangle$
indicates $\tilde{\Upsilon}_{\mathsf{em}}'$ is just the total dephasing
rate for the oscillation $|\mathsf{e}\rangle\leftrightarrow|\mathsf{m}\rangle$.
In the steady state, their steady state values give 
\begin{equation}
\frac{\langle\hat{\text{\textsc{n}}}_{\mathsf{a}}\rangle_{\text{ss}}}{\langle\hat{\text{\textsc{n}}}_{\mathsf{m}}\rangle_{\text{ss}}}=\frac{\Gamma_{\mathsf{ma}}^{-}}{\Gamma_{\mathsf{ma}}^{+}}=e^{\Omega_{\mathsf{ma}}/T},\qquad\frac{\langle\hat{\text{\textsc{n}}}_{\mathsf{b}}\rangle_{\text{ss}}}{\langle\hat{\text{\textsc{n}}}_{\mathsf{e}}\rangle_{\text{ss}}}=\frac{\Gamma_{\mathsf{eb}}^{-}}{\Gamma_{\mathsf{eb}}^{+}}=e^{\Omega_{\mathsf{eb}}/T},\qquad\frac{\langle\hat{\text{\textsc{n}}}_{\mathsf{m}}\rangle_{\text{ss}}}{\langle\hat{\text{\textsc{n}}}_{\mathsf{e}}\rangle_{\text{ss}}}=\frac{\Gamma_{\mathsf{em}}^{-}+\tilde{\Omega}_{\text{d}}^{2}/2\tilde{\Upsilon}_{\mathsf{em}}'}{\Gamma_{\mathsf{em}}^{+}+\tilde{\Omega}_{\text{d}}^{2}/2\tilde{\Upsilon}_{\mathsf{em}}'}.\label{eq:n4-res}
\end{equation}
 Together with $\langle\hat{\text{\textsc{n}}}_{\mathsf{e}}\rangle+\langle\hat{\text{\textsc{n}}}_{\mathsf{m}}\rangle+\langle\hat{\text{\textsc{n}}}_{\mathsf{a}}\rangle+\langle\hat{\text{\textsc{n}}}_{\mathsf{b}}\rangle=1$,
their specific values can be obtained. When $\tilde{\Omega}_{\text{d}}\rightarrow\infty$,
we have $\langle\hat{\text{\textsc{n}}}_{\mathsf{a}}\rangle_{\text{ss}}/\langle\hat{\text{\textsc{n}}}_{\mathsf{b}}\rangle_{\text{ss}}\rightarrow e^{(\Omega_{\mathsf{ma}}-\Omega_{\mathsf{eb}})/T}$.
Thus, if $\Omega_{\mathsf{em}}$ is in the optical frequency regime,
under room temperature, the population ratio $\langle\hat{\text{\textsc{n}}}_{\mathsf{a}}\rangle_{\text{ss}}/\langle\hat{\text{\textsc{n}}}_{\mathsf{b}}\rangle_{\text{ss}}$
is almost zero, namely, $\langle\hat{\text{\textsc{n}}}_{\mathsf{b}}\rangle_{\text{ss}}\simeq1$,
$\langle\hat{\text{\textsc{n}}}_{\mathsf{a,m,e}}\rangle_{\text{ss}}\simeq0$.

To calculate the cooling and heating rates $A_{\pm}$ {[}Eq.\,(\ref{eq:A+-}){]},
we need the equations of $\langle\hat{\sigma}^{\pm}(t)\rangle$, i.e.,
\begin{equation}
\partial_{t}\langle\hat{\sigma}^{+}\rangle=-\frac{1}{2}(\Gamma_{\mathsf{eb}}^{+}+\Gamma_{\mathsf{ma}}^{+}+\tilde{\gamma}_{\text{p}}^{\mathsf{a}}+\tilde{\gamma}_{\text{p}}^{\mathsf{b}})\langle\hat{\sigma}^{+}\rangle:=-\tilde{\Upsilon}_{\mathsf{ab}}'\langle\hat{\sigma}^{+}\rangle.\label{eq:Teb-4L}
\end{equation}
Clearly, here $\tilde{\Upsilon}_{\mathsf{ab}}':=\frac{1}{2}(\Gamma_{\mathsf{eb}}^{+}+\Gamma_{\mathsf{ma}}^{+}+\tilde{\gamma}_{\text{p}}^{\mathsf{a}}+\tilde{\gamma}_{\text{p}}^{\mathsf{b}})$
is the total dephasing rate for the oscillation $|\mathsf{a}\rangle\leftrightarrow|\mathsf{b}\rangle$.
It is worth noting that, unlike the three level system situation {[}Eq.\,(\ref{eq:Teb}){]},
here the equation of $\langle\hat{\sigma}^{+}\rangle$ is no longer
coupled with the other dynamical variables. According to the quantum
regression theorem, the correlation function $\langle\hat{\sigma}^{+}(t)\hat{\sigma}^{-}(0)\rangle_{\text{ss}}$
follows the same equation form as that of $\langle\hat{\sigma}^{+}(t)\rangle$
{[}Eq.\,(\ref{eq:Teb-4L}){]}. As a result, similarly as the discussions
around Eq.\,(\ref{eq:Vt=00003DG.V0}), here the heating and cooling
rates are obtained as 
\begin{align}
A_{+}' & =\frac{2g^{2}}{\tilde{\Upsilon}_{\mathsf{ab}}'}\langle\hat{\text{\textsc{n}}}_{\mathsf{a}}\rangle_{\text{ss}}=\frac{2g^{2}}{\tilde{\Upsilon}_{\mathsf{ab}}'}\frac{e^{\frac{\Omega_{\mathsf{ma}}}{T}}(\Gamma_{\mathsf{em}}^{-}+\tilde{\Omega}_{\text{d}}^{2}/2\tilde{\Upsilon}_{\mathsf{em}}')}{(1+e^{\frac{\Omega_{\mathsf{ma}}}{T}})\Gamma_{\mathsf{em}}^{-}+(1+e^{\frac{\Omega_{\mathsf{eb}}}{T}})\Gamma_{\mathsf{em}}^{+}+(2+e^{\frac{\Omega_{\mathsf{eb}}}{T}}+e^{\frac{\Omega_{\mathsf{ma}}}{T}})\,\tilde{\Omega}_{\text{d}}^{2}/2\tilde{\Upsilon}_{\mathsf{em}}'},\nonumber \\
A_{-}' & =\frac{2g^{2}}{\tilde{\Upsilon}_{\mathsf{ab}}'}\langle\hat{\text{\textsc{n}}}_{\mathsf{b}}\rangle_{\text{ss}}=\frac{2g^{2}}{\tilde{\Upsilon}_{\mathsf{ab}}'}\frac{e^{\frac{\Omega_{\mathsf{eb}}}{T}}(\Gamma_{\mathsf{em}}^{+}+\tilde{\Omega}_{\text{d}}^{2}/2\tilde{\Upsilon}_{\mathsf{em}}')}{(1+e^{\frac{\Omega_{\mathsf{ma}}}{T}})\Gamma_{\mathsf{em}}^{-}+(1+e^{\frac{\Omega_{\mathsf{eb}}}{T}})\Gamma_{\mathsf{em}}^{+}+(2+e^{\frac{\Omega_{\mathsf{eb}}}{T}}+e^{\frac{\Omega_{\mathsf{ma}}}{T}})\,\tilde{\Omega}_{\text{d}}^{2}/2\tilde{\Upsilon}_{\mathsf{em}}'}.\label{eq:A+-1}
\end{align}

Here the driving strength $\tilde{\Omega}_{\text{d}}$ no longer appears
in the correction factor $2g^{2}/\tilde{\Upsilon}_{\mathsf{ab}}'$
as the three level system situation {[}Eqs.\,(\ref{eq:A+}, \ref{eq:A-}){]}.
Therefore, with the increase of the driving light intensity, the cooling
(heating) rate here increases (decreases) monotonically. When there
is no driving light, the ratio between the cooling and heating rates
gives $A_{-}'/A_{+}'=\langle\hat{\text{\textsc{n}}}_{\mathsf{b}}\rangle_{\text{ss}}/\langle\hat{\text{\textsc{n}}}_{\mathsf{a}}\rangle_{\text{ss}}=e^{\omega_{\text{\textsc{r}}}/T}$
{[}see from Eq.\,(\ref{eq:n4-res}){]}, which naturally returns to
the Boltzmann ratio. When the driving strength $\tilde{\Omega}_{\text{d}}\rightarrow\infty$,
the populations could be fully concentrated in the ground state, i.e.,
$\langle\hat{\text{\textsc{n}}}_{\mathsf{b}}\rangle_{\text{ss}}\rightarrow1$,
$\langle\hat{\text{\textsc{n}}}_{\mathsf{a}}\rangle_{\text{ss}}\rightarrow0$.
In this case, the steady state photon number becomes 
\begin{equation}
\langle\hat{n}\rangle_{\text{ss}}=\frac{A_{+}'+\kappa\bar{\mathsf{n}}_{\text{\textsc{r}}}}{A_{-}'-A_{+}'+\kappa}\stackrel{\tilde{\Omega}_{\text{d}}\rightarrow\infty}{\longrightarrow}\bar{\mathsf{n}}_{\text{\textsc{r}}}\big/\big(\frac{2g_{N}^{2}}{\kappa\tilde{\Upsilon}_{\mathsf{ab}}'}+1\big)\simeq\bar{\mathsf{n}}_{\text{\textsc{r}}}\big/\frac{2g_{N}^{2}}{\kappa\tilde{\Upsilon}_{\mathsf{ab}}'}.
\end{equation}
Here the atom-resonator coupling strength has been modified as $g_{N}=\sqrt{N}g$
for the situation of many atoms. 

\end{widetext} 

\bibliographystyle{apsrev4-2}
\bibliography{Refs}

\end{document}